\documentclass[a4paper,11pt]{article}
\pdfoutput=1

\usepackage{jheppub}

\usepackage[T1]{fontenc}

\usepackage{rotating}
\usepackage{afterpage}

\usepackage{xspace,units,graphicx}

\newcommand{\numu}{\ensuremath{\nu_{\mu}}\xspace}
\newcommand{\numubar}{\ensuremath{\overline{\nu}_{\mu}}\xspace}
\newcommand{\nue}{\ensuremath{\nu_{e}}\xspace}
\newcommand{\nuebar}{\ensuremath{\overline{\nu}_{e}}\xspace}

\title{The Results of MINOS and the Future with MINOS+}

\author[a,1]{A.Timmons,\note{For the MINOS collaboration.}}

\affiliation[a]{University of Manchester,\\Department of Physics and Astronomy, Oxford Road, Manchester, M13 9PL, United Kingdom}

\emailAdd{ashley.timmons@manchester.ac.uk}

\abstract{
The MINOS experiment took data from 2005 up until 2012. This was superseded by MINOS+, the continuation of the two-detector, on-axis, long-baseline experiment based at Fermilab, and at the Soudan Underground Laboratory in northern Minnesota. By searching for the deficit of muon neutrinos at the Far Detector, MINOS/MINOS+ is sensitive to the atmospheric neutrino oscillation parameters $\Delta m^{2}_{32}$ and $\theta_{23}$. By using the full MINOS data set looking at both \numu disappearance and \nue appearance in both neutrino and anti-neutrino configurations at the NuMI beam along with atmospheric neutrino data recorded at the FD, MINOS has made the most precise measurement of $\Delta m^{2}_{32}$. Using a full three-flavour framework and searching for \nue appearance MINOS/MINOS+ gains sensitivity to $\theta_{13}$, the mass hierarchy and the octant of $\theta_{23}$. Exotic phenomenon is also explored with the MINOS detectors looking for non-standard interactions and sterile neutrinos. The current MINOS+ era goals are to build on the previous MINOS results improving the precision on the three-flavour oscillation parameter measurements and strengthening the constraints placed on the sterile neutrino parameter space.
}

\begin{document} 
\maketitle
\flushbottom

\section{Introduction}
Over the last couple of decades physicists across the world have obtained model independent evidence for neutrino oscillations. It was in 1998 that Super-Kamiokade~\cite{ref:SuperK} observed muon neutrinos changing flavour as they transversed the atmosphere. Later in 2001 the Sudbury Neutrino Observatory experiment~\cite{ref:SNO} observed neutrinos oscillating between flavours which originated from the sun. Evidence for reactor anti-neutrinos was seen in 2002 with the KamLAND~\cite{ref:KamLAND} experiment. There are now multiple generations of experiments designed to confirm and probe the nature of neutrino oscillations using a neutrino source from an accelerator~\cite{ref:OPERA,ref:T2K}, solar neutrinos~\cite{ref:SAGE,ref:GALLEX}, and nuclear reactor anti-neutrinos~\cite{ref:Chooz,ref:Reno,ref:DoubleChooz,ref:DayaBay}. The implication of neutrino flavour change is indicative that the neutrino must have a non zero mass and violate lepton number conservation; a clear observation of new physics beyond the standard model. 

It was during this time of discovery that the MINOS~\cite{ref:MINOSPro} experiment was proposed. MINOS was designed with a long-baseline and two detectors 1.04~km and 735~km from the neutrino production target respectively; with the goal to measure the atmospheric neutrino oscillation parameters. A previously constructed long-baseline, two detector experiment K2K~\cite{ref:K2K} in Japan had the same goal. MINOS was unique in that its magnetised detectors allowed one to distinguish \numu and \numubar interactions on an event by event basis. The MINOS detectors encountered a higher flux of neutrinos than K2K and over the next few years MINOS contributed to the the era of precision measurements of the fundamental parameters governing this quantum mechanical effect of neutrino oscillations. The MINOS experiment helped show how effectively a two-detector experiment can minimise the large systematic uncertainties associated with a neutrino interaction experiment. In 2012 the MINOS experiment ended and it is the continuation of the detectors taking data in the upgraded accelerator for the NO$\nu$A~\cite{ref:Nova} experiment that the MINOS+~\cite{ref:MINOSPlusProposal} experiment was born. MINOS+ began taking data in September 2013, with a higher flux of neutrinos at high energies MINOS+ becomes sensitive to beyond the standard model neutrinos physics such as sterile neutrinos and large extra dimensions.

The theory of neutrino oscillation describes the change in neutrino flavour composition seen in data. These oscillations arise due to a mixture between mass and flavour eigenstates; three active flavours of neutrino $\left(\nu_{e}, \nu_{\mu}, \nu_{\tau}\right)$ and three mass eigenstates $\left(\nu_{1}, \nu_{2}, \nu_{3}\right)$ are required to fully describe the neutrino oscillations observed in data. The energy dependance of these oscillations are governed by the difference of the square of the mass eigenstates, $\Delta m^{2}_{32}$ and $\Delta m^{2}_{21}$, while the degree of mixing (the amplitude of the oscillations) is governed by three mixing angles $\theta_{12}, \theta_{13}, \theta_{23}$, and a $CP$ violating phase $\delta_{13}$. These parameters make up a $3\times3$ rotation matrix known as the PMNS rotation matrix~\cite{ref:PMNS}.

The difference between the two mass splittings is almost two orders of magnitude, $\theta_{13}$ has been shown to be small by measurements of this parameter by reactors experiments~\cite{ref:DayaBay,ref:Renoth13,ref:DoubleChoozth13} and so one can approximately decouple the two frequencies into two distinct regimes: the ``solar'' oscillation regime is driven by $\Delta m^{2}_{21}$ and $\theta_{12}$ and mostly determines the flavour composition of $\nu_{e}$ particles propagating from within the sun towards Earth. The ``atmospheric'' oscillation regime is driven primarily by $\Delta m^{2}_{32}$ and $\theta_{23}$ and their values govern neutrino oscillation observed in \numu neutrinos decaying from secondary hadrons due to cosmic rays interacting within the Earth's atmosphere. The MINOS experiment was designed to probe the atmospheric sector by using \numu neutrinos from a man made source. MINOS also has sensitivity to the parameter $\theta_{13}$ through observing \nue and \nuebar appearance, allowing for a full analysis combining both disappearance and appearance data. 

\subsection{Oscillation Physics at MINOS}

The neutrino oscillation regime MINOS is most sensitive to is driven by the larger of the two mass-splitting differences $\Delta m^{2}_{32}$; consequently a two-flavour approximation can be used to describe the data using a single mass splitting $\Delta m^{2}$ and effective mixing angle $\theta$. Using this approximation one can express the muon neutrino survival probability as 

\begin{equation}
P(\numu\rightarrow\numu) = 1 - \sin^{2}(2\theta)\sin^{2}
\left(
\frac{1.27\Delta m^{2}[\textrm{eV}^{2}] L_{\nu}[\textrm{km}]}
{E_{\nu}[\textrm{GeV}]}
\right),\label{eqn:NuMuDisappearance}
\end{equation}

\noindent where $L_{\nu}$ is the neutrino propagation distance and $E_{\nu}$ is the neutrino energy. Previous analyses by MINOS rely on this two-favour approximation. However, the neutrino community is entering an era of precision measurement; the error on the atmospheric mass splitting is down to a few percent level and with the discovery of a non zero $\theta_{13}$ in 2012 by Daya bay~\cite{ref:DayaBay}, $\theta_{13}$ has now become one of the most precisely measured angles, and so the need to move to a fuller treatment of neutrino oscillations is ever present. 

Within a three-flavour framework, the oscillations are driven by the mass splittings $\Delta m^{2}_{32}$ and $\Delta m^{2}_{21}$, where $\Delta m^{2}_{31} = \Delta m^{2}_{32} + \Delta m^{2}_{21}$. For exact calculations of the oscillation probabilities one must consider all parameters due to interferences. One can modify the two-flavour oscillation probability parameters in Eq.(\ref{eqn:NuMuDisappearance}) as follows:

\begin{equation}
\sin^{2}2\theta = 4\sin^{2}\theta_{23}\cos^{2}\theta_{13}\left(1-\sin^{2}\theta_{23}\cos^{2}\theta_{13}\right),
\label{eqn:Approx1}
\end{equation}

\begin{equation}
\Delta m^{2} = \Delta m^{2}_{32} + \Delta m^{2}_{21} \sin^{2}\theta_{12} + \Delta m^{2}_{21} \cos\delta_{CP} \sin\theta_{13}\tan\theta_{23}\sin 2 \theta_{12}.
\label{eqn:Approx2}
\end{equation}

Only by moving to a three-flavour framework can the degeneracies between the octant of $\theta_{23}$ and determination of the mass hierarchy (sign of $|\Delta m^{2}_{32}|$) be broken.
These equations only account for neutrino oscillation within a vacuum, however, when neutrinos traverse through matter the eigenstates become modified due to the MSW effect~\cite{ref:MSW1,ref:MSW2}. To account for this one can replace $\theta_{13}$ with a modified mixing angle $\theta_{M}$, given by~\cite{ref:thetamod}, such that

\begin{equation}
\sin^{2}2\theta_{M} = \frac{\sin^{2}2\theta_{13}}{\sin^{2}2\theta_{13} + \left(A - \cos2\theta_{13}\right)^{2}},
\label{eqn:MatterEffect1}
\end{equation}

\noindent where the magnitude of $A$ determines the size of the matter effect and can be expressed as $A = \pm2\sqrt{2}G_{F}n_{e}E_{\nu}/\Delta m^{2}_{31}$, where $G_{F}$ is the Fermi weak coupling constant and $n_{e}$ is the density of electrons in the medium. The sign of A is positive (negative) for neutrinos (anti-neutrinos). The magnitude of $\sin^{2}2\theta_{M}$ in equation \ref{eqn:MatterEffect1} influences the amount of mixing between $\nu_{\mu} \leftrightarrow \nu_{e}$. This MSW mechanism has an effect on a \numu disappearance analysis through $\nu_{\mu} \leftrightarrow \nu_{e}$ mixing, therefore, in order to perform a precision measurement using \numu disappearance one must also take into account \nue appearance.

MINOS is also sensitive to $\nu_{e}$ interactions, the probability for $\nu_{e}$ appearance to second order can be approximated to~\cite{ref:MatterMod}:

\begin{align}
P \left( \nu_{\mu} \rightarrow \nu_{e} \right)
\approx & \sin^{2}\theta_{23}\sin^{2}2\theta_{13}\frac{\sin^{2}\Delta\left(1-A\right)}{\left(1-A\right)^{2}} \notag \\
&+ \alpha \tilde{J}\cos\left(\Delta \pm \delta_{CP}\right)\frac{\sin \Delta A}{A}\frac{\sin\Delta\left(1-A\right)}{1-A} \notag \\
&+ \alpha^{2}\cos^{2}\theta_{23}\sin^{2}2\theta_{12}\frac{\sin^{2}\Delta A}{A^{2}},
\label{eqn:MatterEffect2}
\end{align}

\noindent where $\alpha \equiv \Delta m^{2}_{21} / \Delta m^{2}_{31} \left( \sim 0.03  \right)$, $\tilde{J} \equiv \cos\theta_{13} \sin2\theta_{13}\sin2\theta_{12}\sin2\theta_{23}$ and  $\Delta \equiv \Delta m^{2}_{31}L_{\nu} / 4 E_{\nu}$. The second term in equation \ref{eqn:MatterEffect2} will have a plus (minus) sign for neutrinos (anti-neutrinos). Being sensitive to \nue and \nuebar appearance allows one to probe the mass hierarchy (as the sign of $\Delta m^{2}_{31}$ will change) as well as the CP violating phase. 

\section{The MINOS experiment}
\subsection{The NuMI Beam}

The Neutrinos at the Main Injector (NuMI) neutrino beam~\cite{ref:NUMI} was built at Fermilab, to provide neutrinos for the MINOS experiment. The NuMI beam typically has a beam power of 350~kW with a design specification of up to 400~kW. A high-intensity beam is required to achieve a meaningful event rate at the MINOS Far Detector (FD) placed several hundred kilometres away. Such a distance significantly reduces the neutrino flux as it falls with the square of the distance from the decay point.

\begin{figure}[!h]
\centering
\includegraphics[trim={0 5cm 0 5cm}, width=\textwidth, clip]{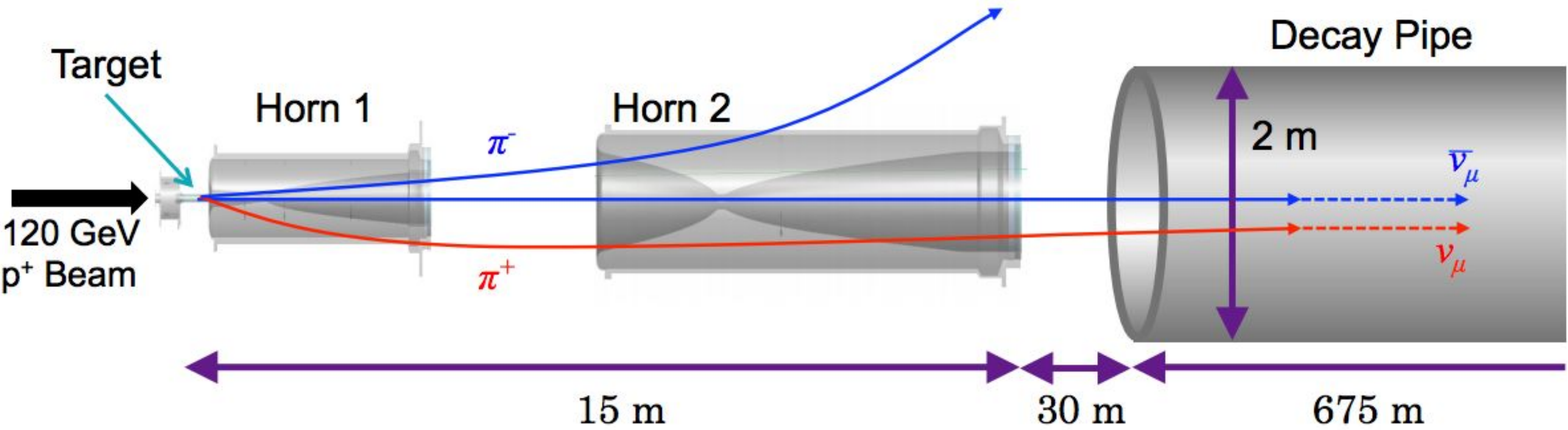}
\caption{The NuMI beam.}
\label{fig:NuMIBeam}
\end{figure}

To produce such a powerful neutrino beam the first stage is to create protons from H\textsuperscript{-} ions. To achieve this the ions are accelerated by a Radio Frequency (RF) quadrupole up to an energy of 750~keV. From there a linear accelerator then accelerates the ions to an energy of 400~MeV which have subsequently passed through a thin carbon foil stripping the electrons off the ions to leave a beam of protons. The protons are fed into a rapid cycling synchrotron (Booster) and accelerated in batches up to energies of 8~GeV. Subsequently they are fed into the Main Injector where they are accelerated to 120~GeV. The Main Injector has a circumference seven times larger than the Booster and so up to six batches can be inserted into the Main Injector at once. It is the interaction between these high energy protons upon a fixed graphite target which results in plethora of charged hadrons (predominantly pions, with a significant kaon component at higher energies). These charged hadrons pass through two parabolic, magnetic horns which focus either positive or negative hadrons depending on the direction of the electric current being pulsed through the horns. The focused hadrons travel along a 675~m decay pipe. Its the decay of these hadrons within the pipe that form the predominately muon flavour neutrino beam. By focusing the positive hadrons a beam of predominately \numu is created (\numu-dominated beam mode), by focusing the negativley charged hadrons the \numubar component can be increased (\numubar-enhanced beam mode) . Figure~\ref{fig:NuMIBeam} shows a diagram of the charged hadrons being focused by the two horns into the decay pipe. 

\begin{figure}[!ht]
\centering
\includegraphics[height=0.50\textwidth]{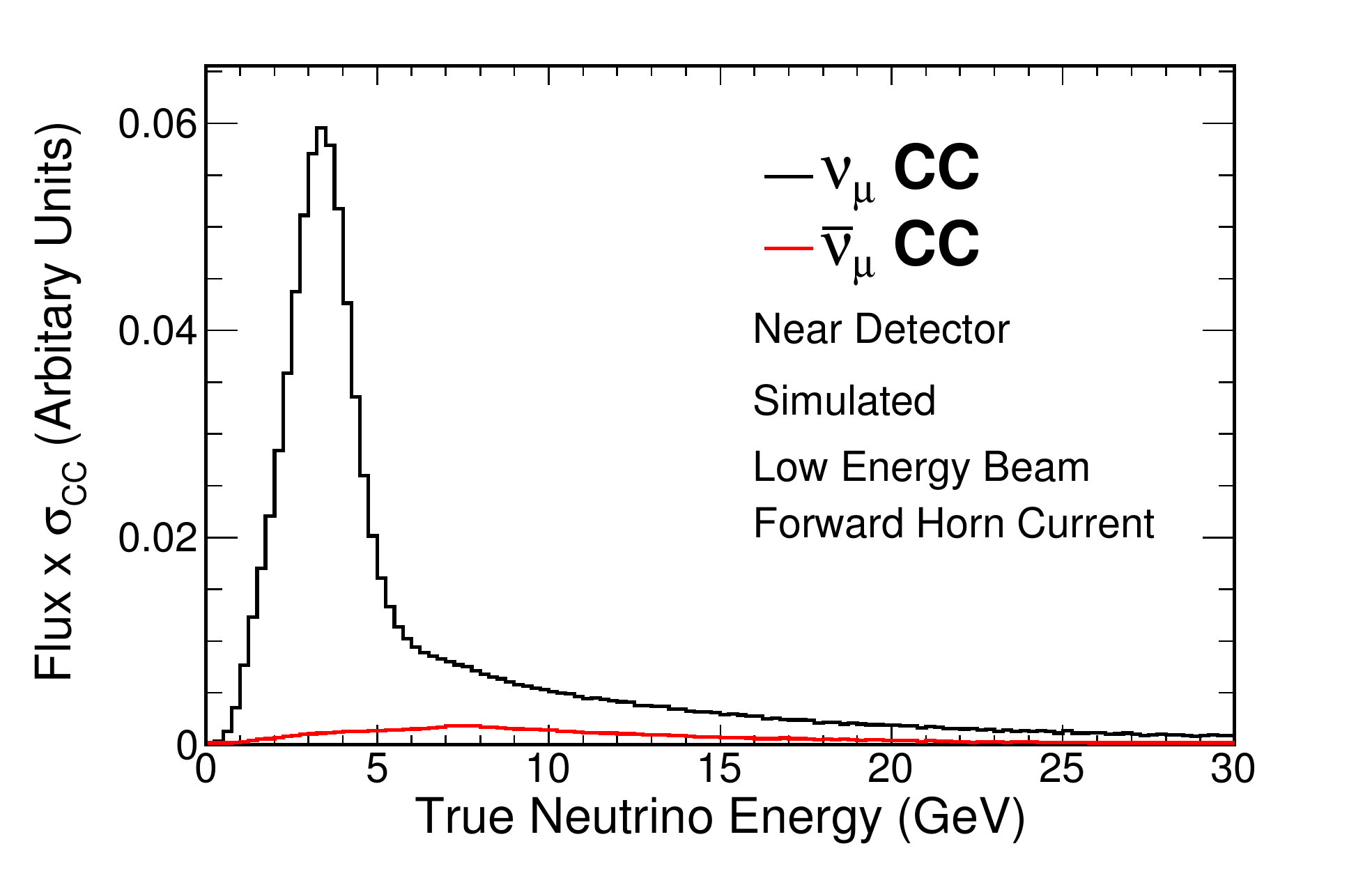}
\includegraphics[height=0.50\textwidth]{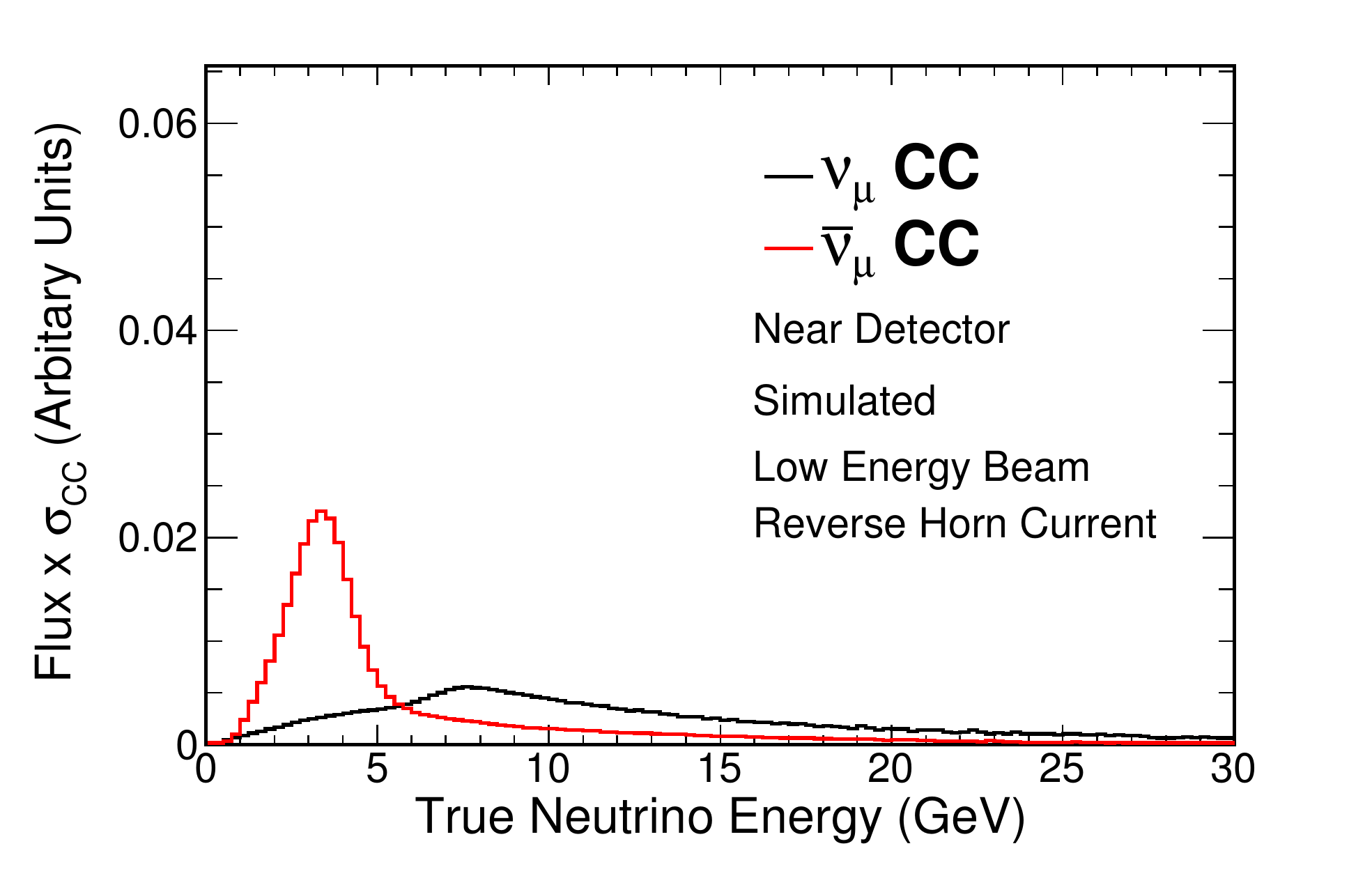}
\caption{The reconstructed neutrino energy spectra at the MINOS Near Detector. The top picture shows the energy spectrum for positively focused hadrons producing a predominately \numu beam. Below is when negatively charged hadron are focused which increases the amount of \numubar events seen at the detectors. Note how in anti-neutrino mode the event rate is significantly less due to different cross sections between neutrinos and anti-neutrinos. }
\label{fig:BeamCompo}
\end{figure}

Figure~\ref{fig:BeamCompo} shows the composition of the NuMI beam for charged current neutrino interactions observed in the MINOS Near Detector. The significant difference in composition and event rate between these beam modes arises mainly from the fact that the \numubar interaction cross section is a factor of approximately two lower than the \numu interaction cross section.

\begin{figure}[!h]
\centering
\includegraphics[trim={0 0 0 0},width=0.8\textwidth,clip]{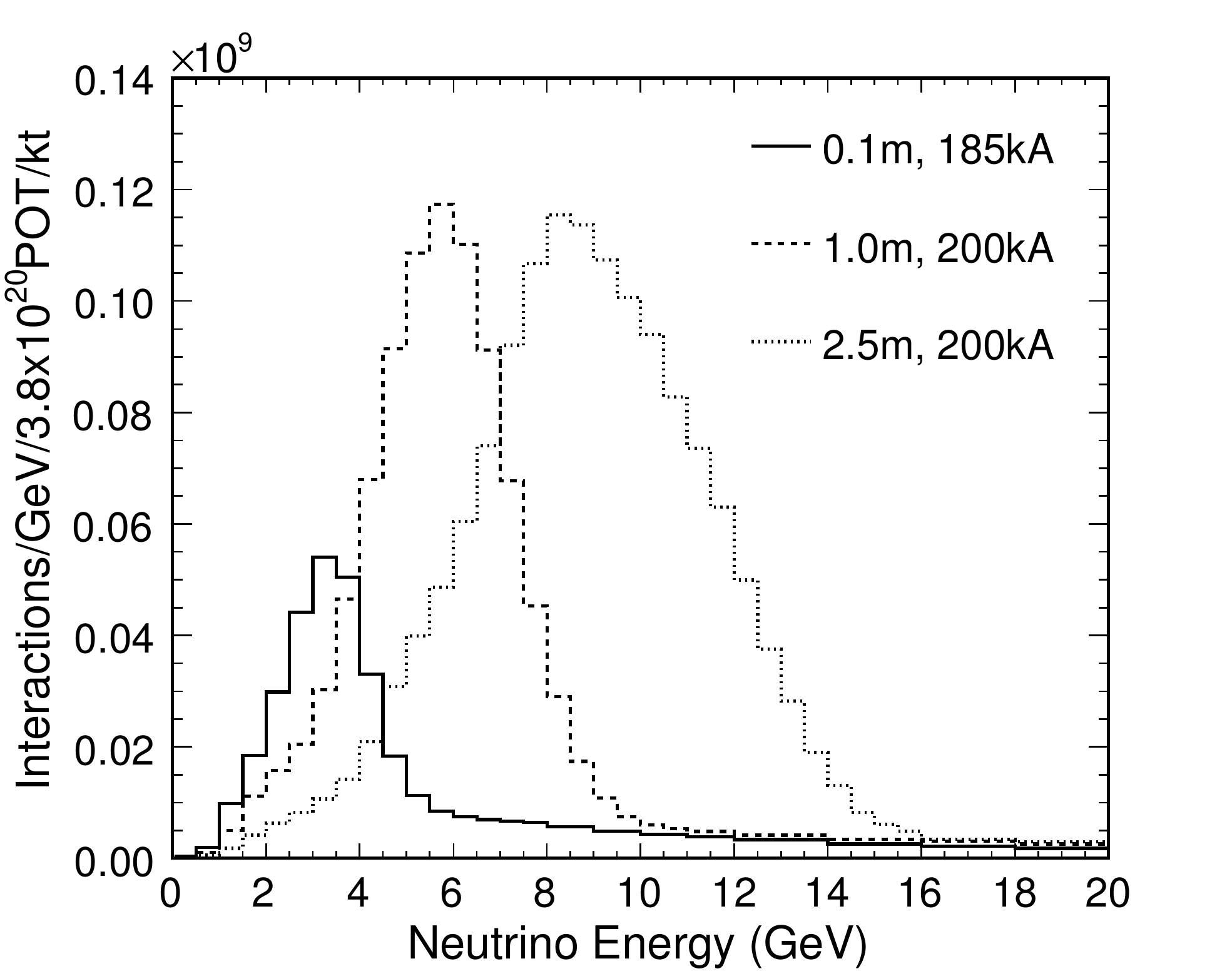}
\caption{The true energy distribution of near detector events from the three different NuMI configurations. The target position (distance upstream of a nominal position) and horn current of each configuration are shown in the caption. ``0.1~m, 185~kA" is the ``low-energy" configuration in which most of the MINOS data has been taken. ``1.0~m, 200~kA" is the ``medium-energy" configuration which is the beam configuration in the NO$\nu$A era and therefore is the beam setup in which MINOS+ is currently taking data in. The final configuration is the ``pseudo-high-energy" configuration.}
\label{fig:NuMIBeamConfig}
\end{figure}

The neutrino energy spectrum provided by the NuMI beam is tunable, through changing the relative positions between the target and the focusing horns. Three of the possible configurations are shown in figure~\ref{fig:NuMIBeamConfig}. The main goal of MINOS was to measure the atmospheric oscillation parameters, this would require a large flux at the oscillation dip which would be located around the 2 GeV region. In the MINOS era the configuration of the NuMI was set to the low energy setting. For the MINOS+ era the NO$\nu$A experiment is $14$mrad off axis and requires the NuMI beam to be set at the medium energy configuration. MINOS/MINOS+ is an on axis experiment (the detectors line up with the beam axis) and therefore MINOS+ observes a high flux of neutrinos at higher energies compared to the low energy configuration.

\afterpage{\clearpage}
\begin{sidewaysfigure}
\includegraphics[scale=1.15]{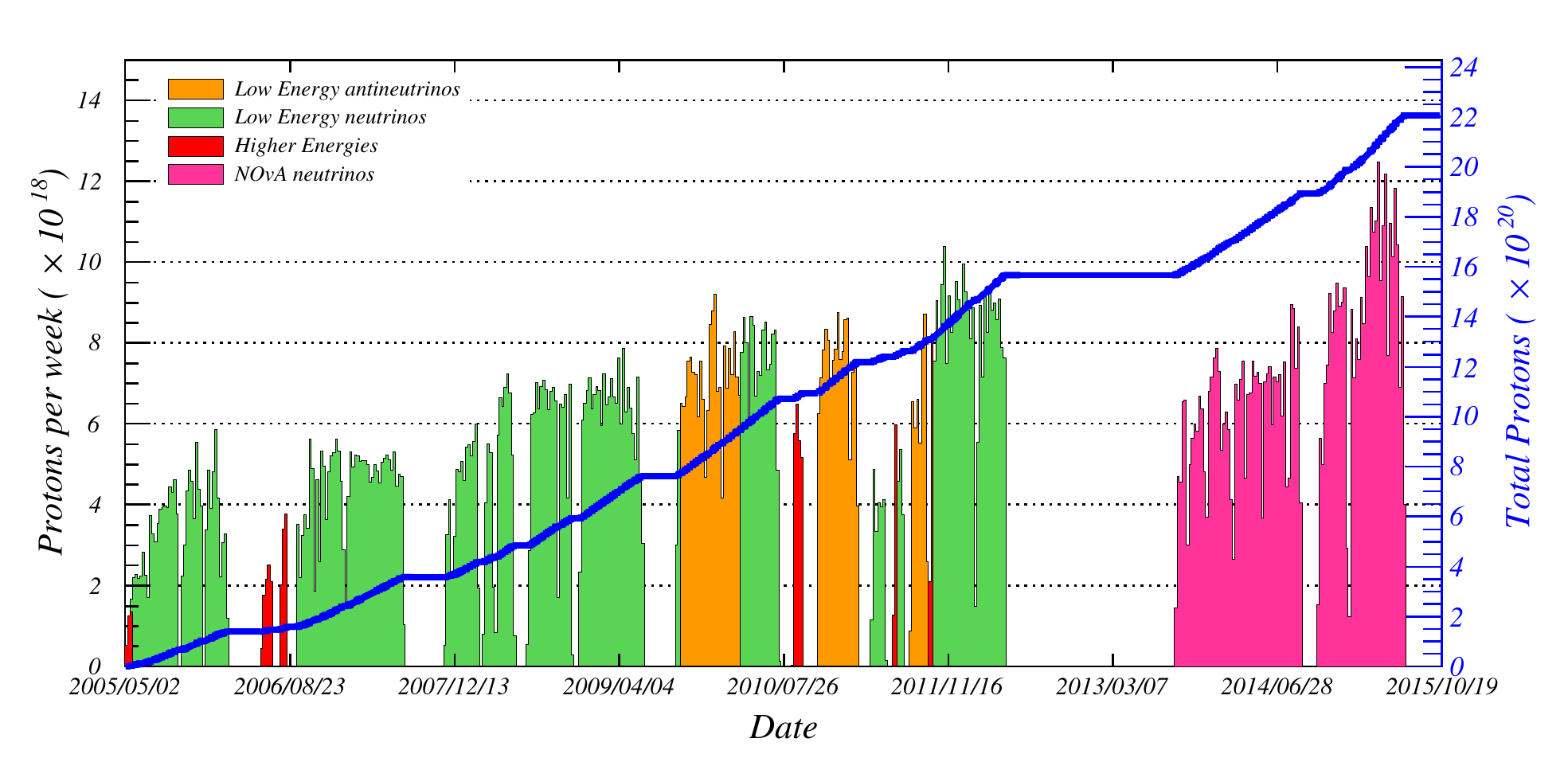}
\caption{Showing the beam configuration for the MINOS and MINOS+ experiments. Most of the data for MINOS was taken in the low energy \numu-beam (in green) and \numubar (orange). Special runs where the beam was configured to higher energies or the magnetic horn turn off is indicated by the red. The Magenta coloured runs are the beginning of the MINOS+ data taking in the \numu-beam during the NO$\nu$A era. }
\label{fig:RunningMode}
\end{sidewaysfigure}

\subsection{The MINOS Detectors}
\begin{figure}[!ht]
\centering
\includegraphics[height=0.30\textwidth]{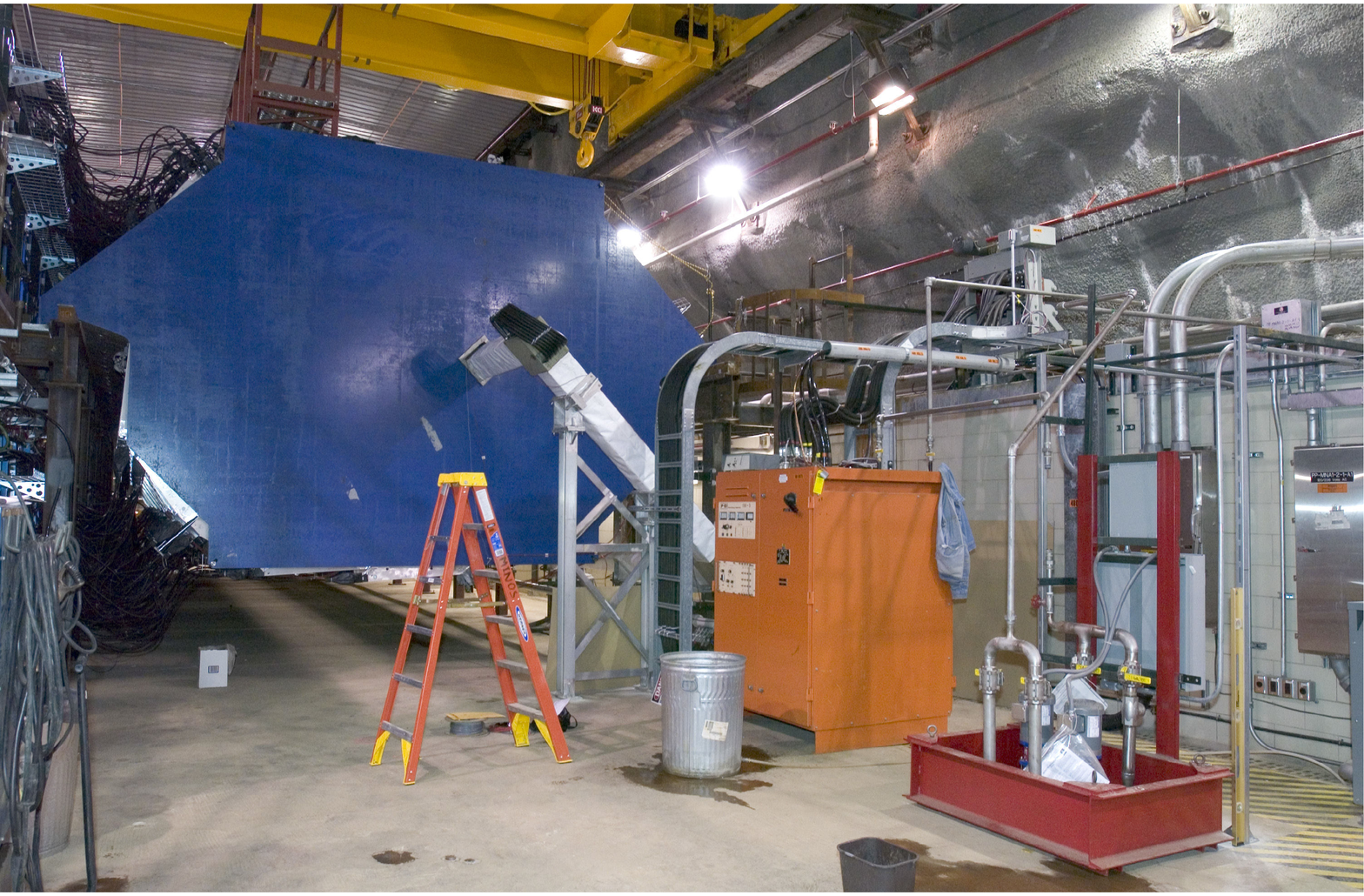}
\includegraphics[height=0.30\textwidth]{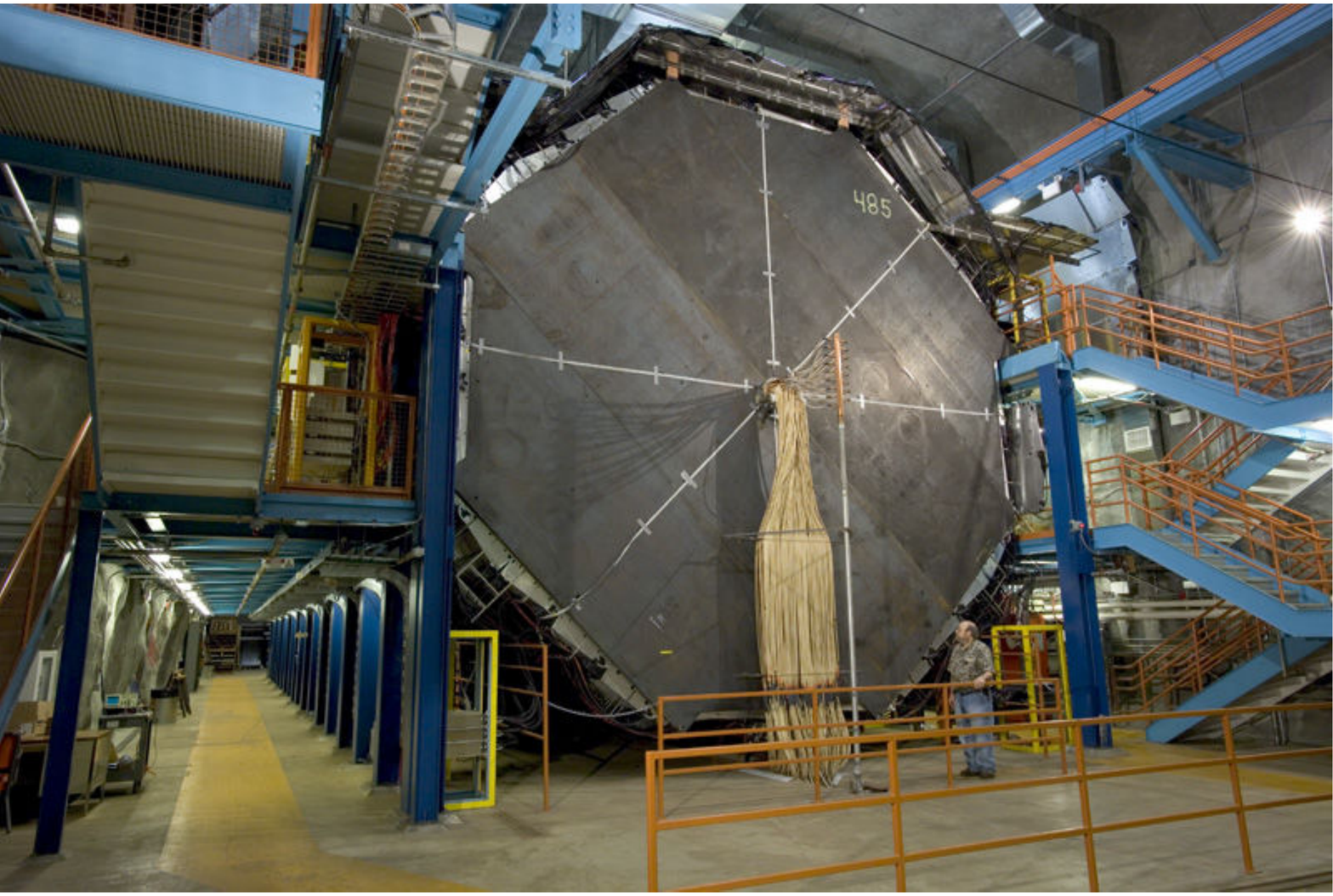}
\caption{The MINOS detectors. Left: the Near Detector at Fermilab; right: the Far Detector at the Soudan Underground Laboratory.}
\label{fig:MINOSDetectors}
\end{figure}

The MINOS experiment has two steel-scintillator calorimeters~\cite{ref:MINOSNIM} designed with the same materials and to operate in an identical manner, known as functionally-identical.  The calorimeters measure the energy deposition and event topologies of neutrino interaction events. The detectors are shown in figure~\ref{fig:MINOSDetectors}. Both detectors are made of alternating layers of 1.00~cm thick plastic-scintillator and 2.54~cm thick steel planes. As neutrinos travel through the detector they interact with the iron-nuclei, and the charged final-state particles travel through the scintillator depositing energy which is read out as light.

\begin{figure}[!ht]
\centering
\includegraphics[height=0.60\textwidth]{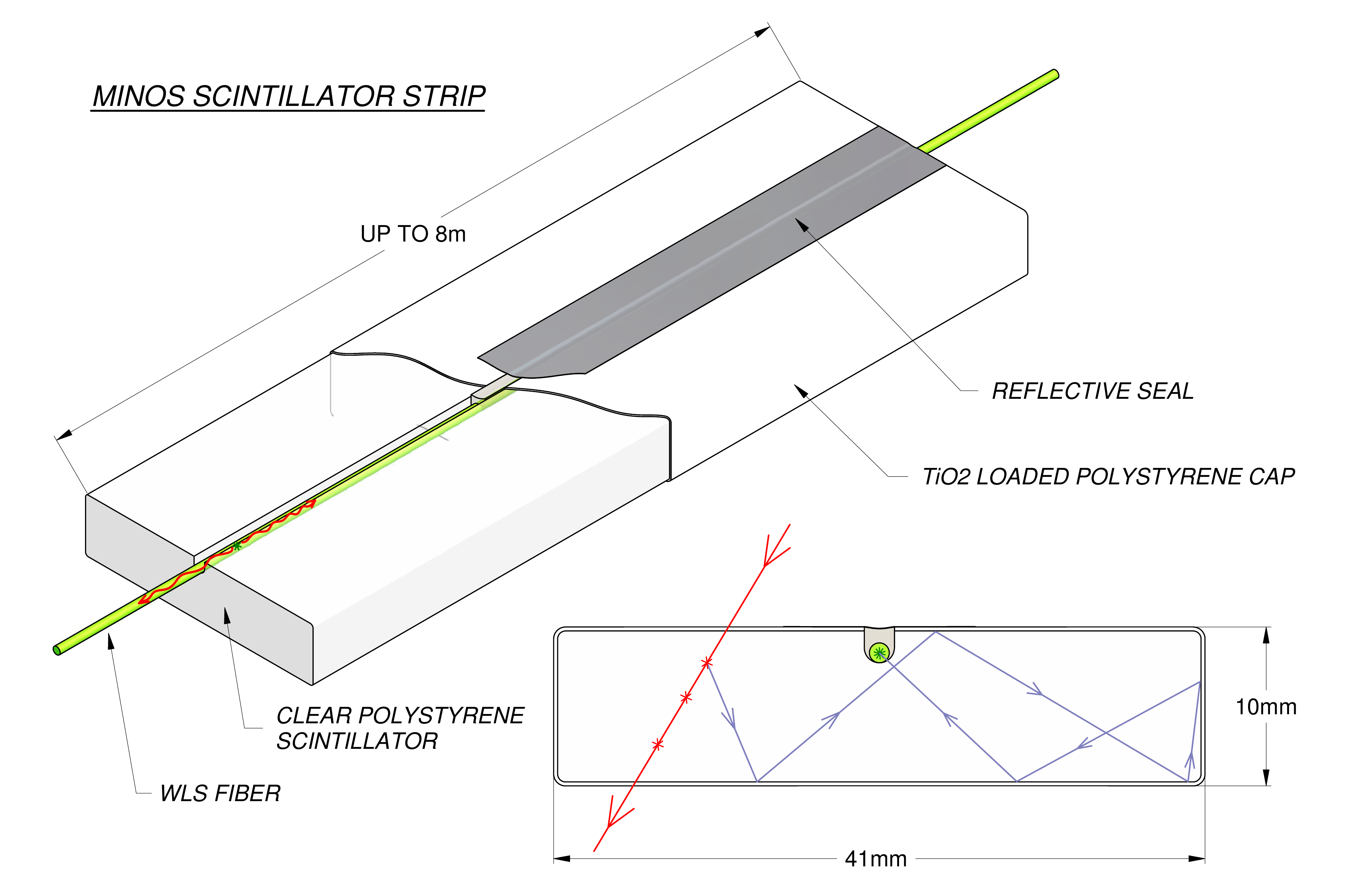}
\caption{A strip of MINOS scintillator with the WLS fiber installed along the center. The image in the lower right corner depicts a charged particle (red line) depositing energy by producing light (blue line) as it passes through the strip. The light travels along the WLS fiber and is read out by a PMT.}
\label{fig:MINOSScintillators}
\end{figure}

The light travels along WaveLength Shifting (WLS) fiber and is read out by a series of a PhotoMultiplier Tubes (PMTs). Figure~\ref{fig:MINOSScintillators} shows a strip of the scintillator used in the MINOS detectors, along the middle a groove is made so that a wavelength shifting fiber can be installed. It is from the light patterns that a neutrino candidate event can be reconstructed, so that information about the topology of the event can be extracted. The steel planes are magnetised by a coil aligned to the longitudinal axes of each detector giving a magnetic field of  approximately 3~T. The trajectories of the charged particles are therefore curved (the direction depends on the polarity of the current in the coil) and thus \numu and \numubar CC interactions can be distinguished.

The Near Detector (ND) is situated 1.04~km downstream from the neutrino target at Fermilab. With a mass of 0.98~kton the ND measures the reconstructed neutrino energy spectrum before oscillations have occurred. The ND has two distinct sections so that it can take advantage of the
high neutrino flux at this location to define a relatively small target fiducial volume for selection of events for the near/far comparison. The section closest to the target is used to define the interaction vertex and measure the energy of the neutrino-induced hadronic shower; every plate is instrumented with plastic scintillator to act as a calorimeter. The second section is used as a muon spectrometer to measure the momenta of energetic muons where one in every five plates is instrumented with scintillator. The scintillator planes are made up of 4 cm-wide strips. The strips on adjacent planes are oriented perpendicular to each other to allow three-dimensional reconstruction of events. The planes are oriented $45^{o}$ to the vertical defining a co-ordinate system referred to as the $u$ and $v$ directions. 

The Far Detector is 735~km downstream from the neutrino production target, 705~m underground in a mineshaft in northern Minnesota. The FD is significantly larger than the ND to compensate for the decrease in the neutrino flux. With a mass of 5.4~kton the FD measures the reconstructed neutrino energy spectrum and will observe a different neutrino flavour composition of the beam due to neutrino oscillation. The geometry is similar to that of the ND in that it is spilt into two ``super modules" of 239 and 247 planes, however the FD does not have a vertex and spectrometer section due to its distance away from the beam, it observes significantly less neutrino interactions. A veto shield composed of layers of scintillator covers the top and sides of the FD to better identify incoming cosmic ray muons that may enter the fiducial volume of the detector helping to obtain a high pure sample of downward-going atmospheric \numu events.

The two detector method is a very powerful experimental setup, since it allows the cancellation of large uncertainties that beset any neutrino oscillation experiment. The uncertainty in neutrino flux and cross sections are only known to tens of percent. Therefore, by looking at the  disappearance and appearance in the FD relative to the ND the uncertainties can be reduced significantly as only the relative uncertainty between the two detectors will affect the final measured result.

\section{Neutrino Interactions in the MINOS Detectors}
\begin{figure}[!ht]
\centering
\includegraphics[trim={0 0cm 0 2cm}, height=0.50\textwidth, clip]{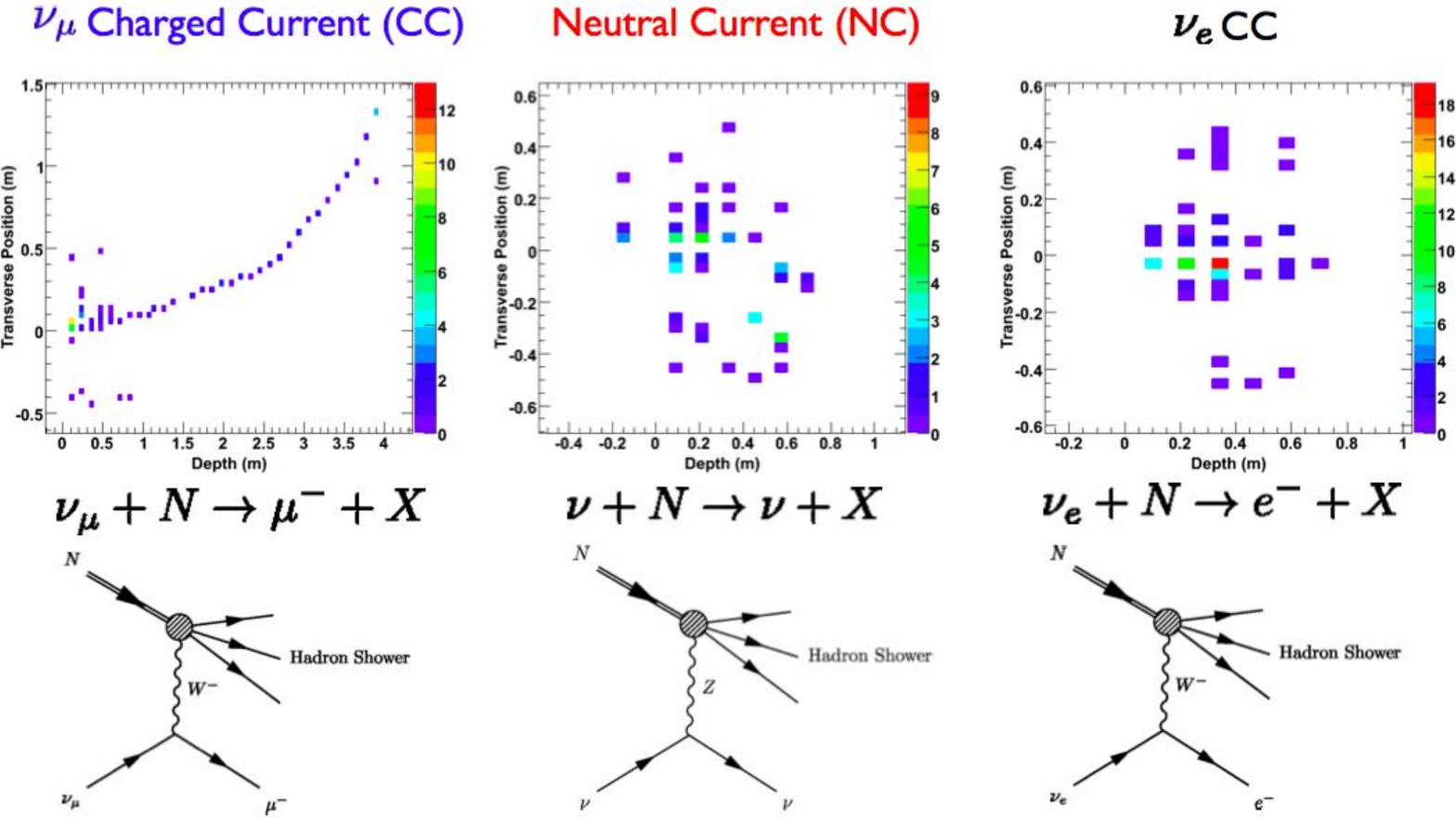}
\caption{Neutrino interaction topologies observed in the MINOS detectors. Left: A charged current \numu interaction. Middle: A neutral current interaction. Right: A charged current \nue
interaction. Each coloured pixel represents a scintillator strip with energy deposited from a charged particle. The colour scale displays the amount of light: purple and blue are low light levels, through to orange and red for the highest light levels.}
\label{fig:MINOSInteractions}
\end{figure}

There are three neutrino interactions that are of interest to MINOS as shown in figure~\ref{fig:MINOSInteractions}.

The main channel is the charged current (CC) \numu (\numubar) interaction
 \[
\numu(\numubar)+X\rightarrow\mu^{-(+)}+X^{\prime}.
\]
The cascade of hadrons, $X^{\prime}$, produces a diffuse shower of energy deposits near the interaction vertex. MINOS was constructed with steel planes so that it can contain a signficant proportion of the final state muons. A muon produces a long track that curves due to the magnetic field. It is the direction of curvature that allows  MINOS to identify the incoming neutrino as a \numu or a \numubar. 

All active neutrino flavours undergo the neutral current (NC) interaction through the process
\[
\nu+X\rightarrow\nu+X^{\prime}.
\]

\noindent Only the hadronic shower is observed, producing a diffuse pattern of energy deposits within the detector. It is not possible to determine the flavour of neutrino.

Finally, electron neutrinos undergo CC interactions through the process
\[
\nu_{e}(\nuebar)+X\rightarrow e^{-(+)}+X^{\prime}.
\]

\noindent The electron gives rise to an electromagnetic shower, which produces a much denser, more compact shower of energy deposits. This interaction is difficult to detect due to the steel plate thickness, meaning only a few events leave any energy deposits in the plastic-scintillator. 

Above a few GeV the dominant process is deep inelastic scattering (DIS). Here the neutrino has sufficient energy that it can resolve the individual quark constituents of the nucleon which manifests in the creation of a hadronic shower. However, at MINOS the oscillation dip observed in muon neutrino disappearance at the far detector occurs just below $2$ GeV;  these neutrino interactions provide a large source of signal events for a neutrino oscillation analysis. At this energy neutrinos can elastically scatter off an entire nucleon liberating a nucleon (or multiple nucleons) from the target. In the case of charged current scattering this process is referred to as ``quasi-elastic scattering" (CCQE). A detailed review on the current state of neutrino cross sections can be found here~\cite{ref:NuCrossSectionReview}.

The energy of a neutrino event is calculated by summing the shower energy deposits and the muon track energy. The energy resolution of contained muon tracks is 4.6\%~\cite{ref:MuonTrk1}. If a muon track exits either one of MINOS detectors then the curvature of the track is used to calculate the energy. The curvature of a muon track is directly proportional to the ratio of its electric charge to its momentum; at the peak of the neutrino beam at around 3~GeV the resolution is 11\%~\cite{ref:MuonTrk1}.

All three interaction processes can result in a shower of energy deposited in the detectors. The MINOS detectors are too coarse to reliably use shower topology information and so the energy is reconstructed calorimetricaly. The final calorimetric hadronic and electromagnetic shower resolutions are well modelled by simulation and the resolution is parametrised as $56\% / \sqrt{E}$ for hadronic showers and $21\% / \sqrt{E} $ for electromagnetic showers, where E is the particle energy in GeV~\cite{ref:MikeThesis,ref:TriciaThesis}.

\begin{figure}[!ht]
\centering
\includegraphics[height=0.70\textwidth]{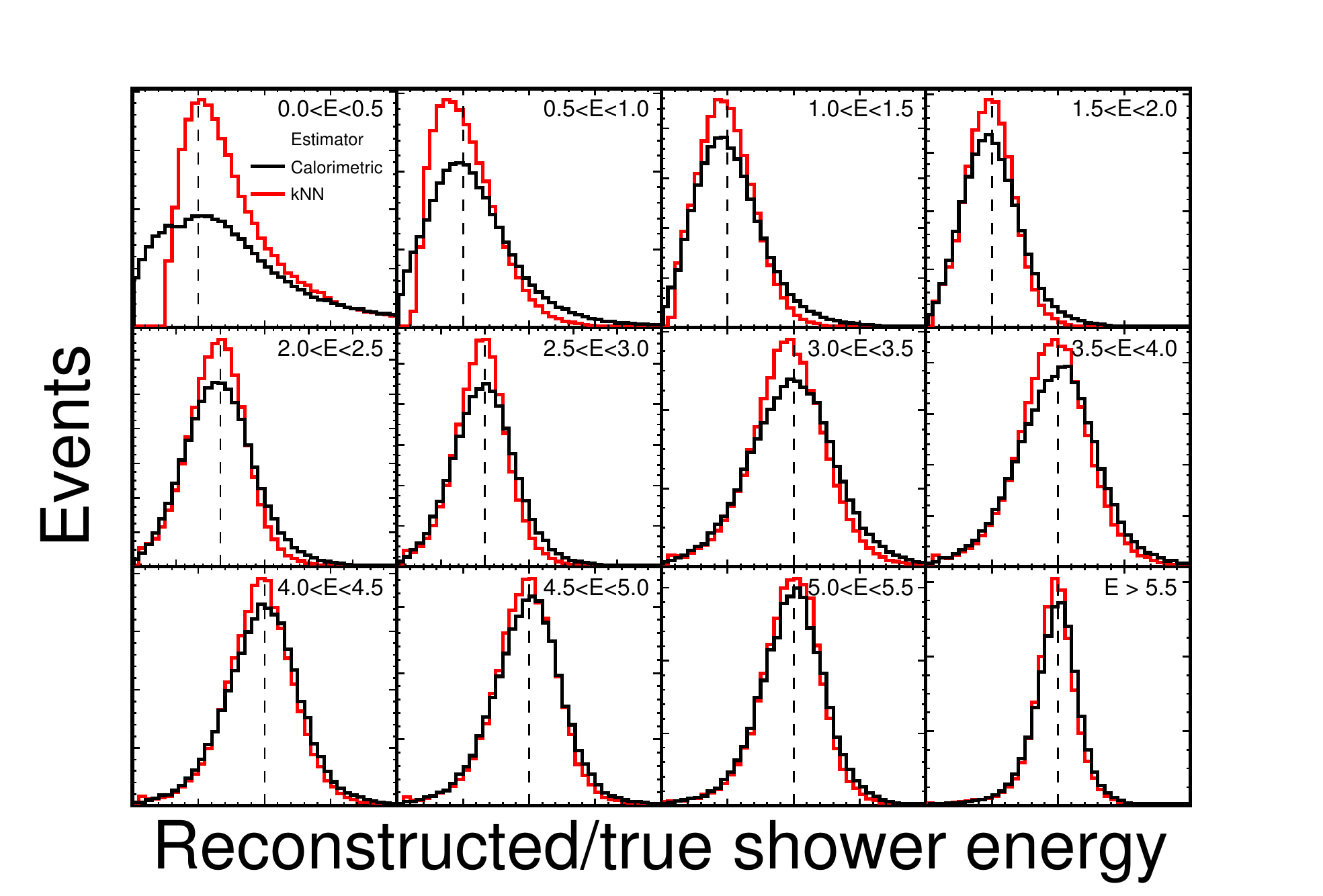}
\caption{ Distribution of reconstructed over true shower energy for different ranges of true shower energy (indicated in the corner of each panel). The calorimetric estimator is shown in black, and the $k$-nearest-neighbour estimator in red. In each panel, the dashed line indicates the position of $E_{\text{reco}}/E_{\text{true}} = 1$. Figure taken from reference~\cite{ref:BackhouseThesis}.}
\label{fig:KNNest}
\end{figure}

A calorimeter response is different for hadronic and electromagnetic showers. A fraction of the energy deposited by the showering particle is invisible, i.e. it does not contribute to the calorimeter signal, this can cause some undesirable properties in a calorimeter causing non-linearities. For CC \numu (\numubar) interactions a more sophisticated method is implemented to measure shower energies~\cite{ref:BackhouseThesis}. A $k$-nearest-neighbour algorithm~\cite{ref:KNNref} uses a multivariate analysis of a broader range of event-level information to provide an estimate of shower energy. The variables are by the algorithm: the event length, the average energy deposited per scintillator plane along the track, the transverse energy deposition, and the fluctuation of the energy deposition along the track.  The $k$-nearest-neighbour algorithm uses the 400 nearest neighbours and comparing them to a simulated library of events the total shower energy of a event can be estimated. Figure~\ref{fig:KNNest} shows the distribution of reconstructed over true shower energy for different ranges of true shower hadronic shower energy using calorimetric energy and  $k$-nearest-neighbour algorithm estimator. It can be seen that the $k$-nearest-neighbour algorithm gives a better estimate of the shower energy at lower energies. This improves the hadronic energy resolution of from 55\% to ~43\% for showers between 1.0 - 1.5 GeV.

\subsection{Selection of charged-current \numu and \numubar interactions}
MINOS was designed to measure the neutrino oscillation parameters in the atmospheric region ($\Delta m^{2}_{32}$ and $\theta_{23}$). To achieve this one needs a sample of \numu (\numubar) CC events. Figure \ref{fig:MINOSInteractions} shows that this can be achieved by selecting neutrino candidate events with a muon track. 

There are three main backgrounds that have an effect on signal purity. At low energies, NC interactions can result in a charged hadron producing a track, thus being mistaken for a \numu-CC event with a low energy muon track. Atmospheric $\numu$ events represent a potential source of muon neutrinos distinct from those in the muon beam and hence are not useful when only considering a beam disappearance oscillation analysis. Wrong-sign events can become a background when the the muon charge deduced from curvature is measured incorrectly if one wishes to separate \numu and \numubar samples to test if they oscillate with different probabilities. 

\begin{figure}[!ht]
\centering
\includegraphics[height=0.50\textwidth]{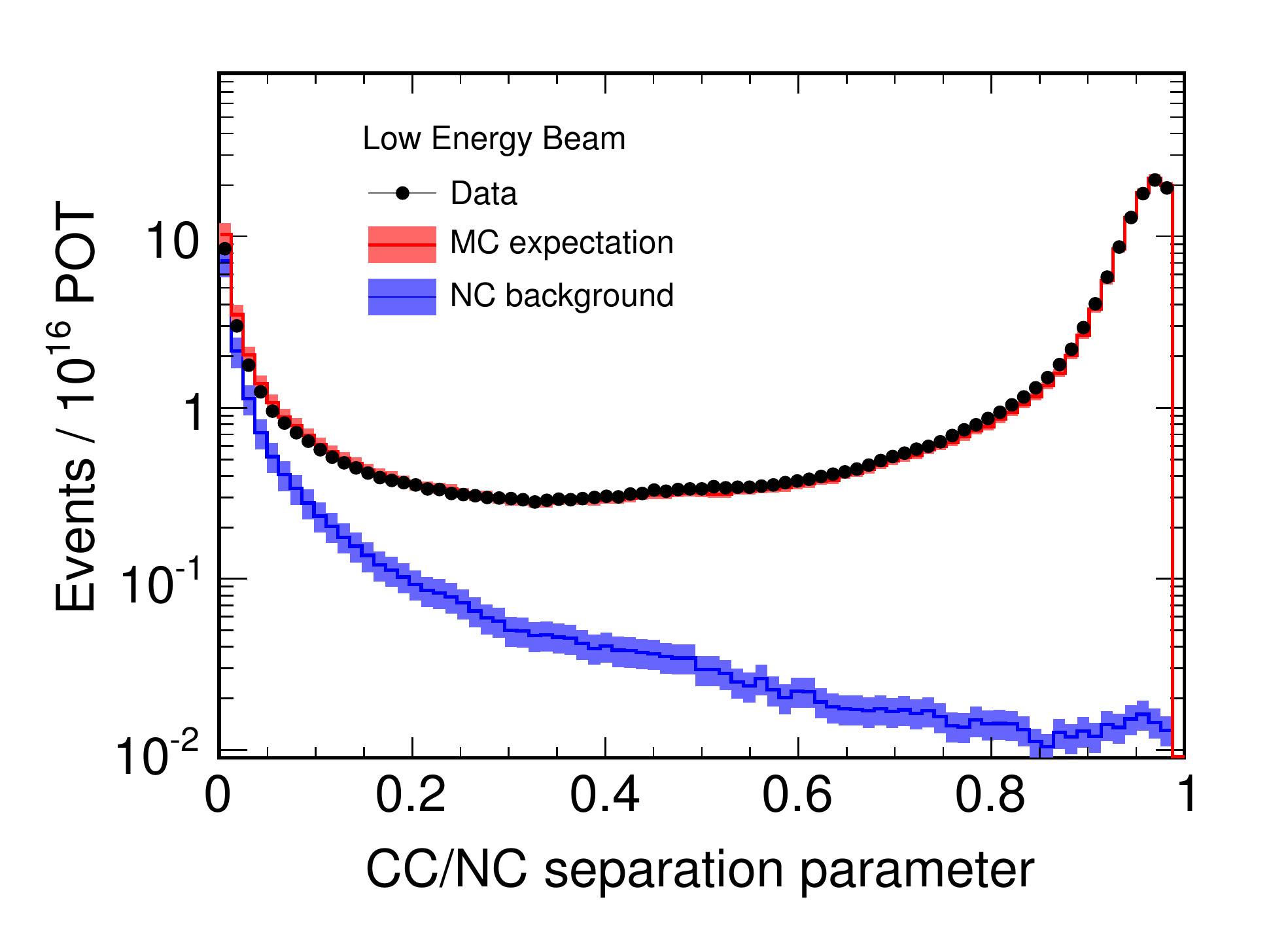}
\caption{The $k$-nearest-neighbour discrimination variable used to separate \numu CC interactions from track-like hadronic backgrounds. Events with a parameter value
greater than 0.3 are selected as \numu CC interactions for analysis.}
\label{fig:KNN}
\end{figure}

To reduce the NC background one needs to separate the NC and CC candidate events. MINOS uses a $k$-nearest-neighbour algorithm; a simulated high statistics data set is created with two known classes of events, one with a muon track and one without. Four variables are used to create a discriminating variable to be applied to all track-like events. The variables are: the number of MINOS detector planes associated with a muon track (muon tracks tend to extend much further than NC showers), the average energy deposited per scintillator plane along the track, the transverse energy deposition profile, and the variation of the energy deposited along the muon track. Figure~\ref{fig:KNN} shows the distribution of the $k$-nearest-neighbour algorithm as a single variable~\cite{ref:RustemThesis}.

The event selection is identical for \numubar and \numu CC events due to their similar interaction topologies. The NuMI beam can be configured to produce an anti-neutrino enhanced beam. From the curvature of the muon track reconstructed by a Kalman Filter~\cite{ref:JohnMarshallThesis} algorithm the lepton number of the neutrino can be deduced. After track identification all the remaining hits which are in proximity to one another are grouped into showers.

\subsection{Selection of charged-current \nue interactions}

The selection for \nue events relies on looking for events with a dense shower arising from electromagnetic interactions from an electron as it passes through the MINOS detectors. The dominant background comes from NC events generating a dense hadronic shower. Such behaviour can be caused by a neutral pion decaying into a pair of photons. The majority of \nue appearance expected at the FD occurs in neutrino events with energy in the range 1-8~GeV and so only this range is considered.

\begin{figure}[!ht]
\centering
\includegraphics[trim={0cm 6cm 0cm 7cm}, height=.65\textwidth, clip]{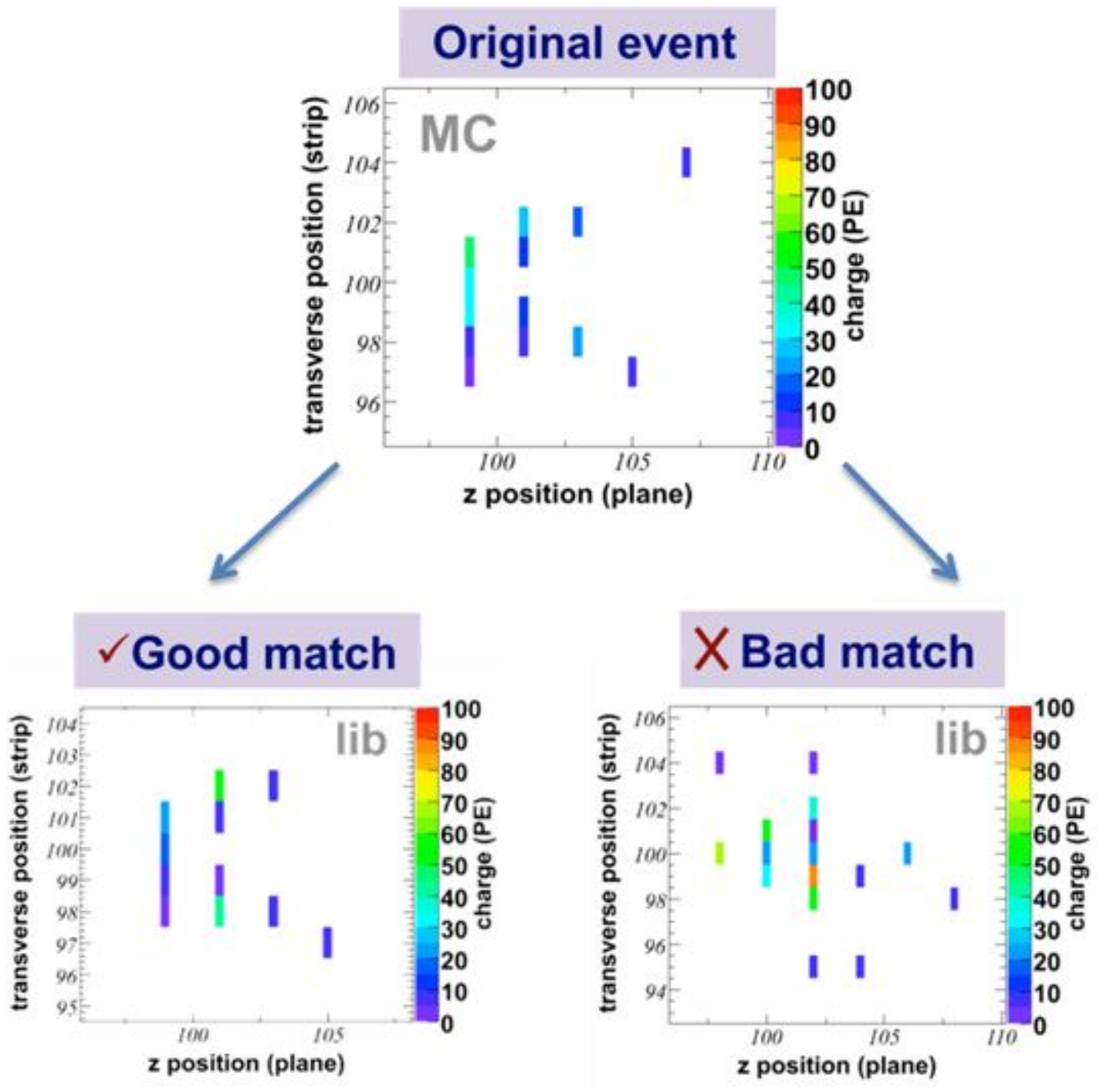}
\caption{Showing an example of a simulated \nue event compared to a good and bad match from the LEM library.}
\label{fig:nuematchevents}
\end{figure}

The granularity of the MINOS detectors makes resolving any topology from electromagnetic showers almost impossible, therefore all candidate events with a shower have their energy deposition patterns compared to a large library of order $10^{7}$ simulated events containing signal (40\%) and background events (60\%). This technique is called Library Event Matching (LEM)~\cite{ref:PedroThesis,ref:RuthThesis,ref:AdamSThesis}. The 50 simulated events that match the event with a similar pattern of energy
deposited in each scintillator strip excited by the shower are chosen. Data and library events are not spatially translated to align them for the best matching. The matching procedure provides a quantitative means of determining the likelihood that two different charge topologies were created by the same primary deposition. For an arbitrary energy deposit, the mean expected charge on a photomultiplier tube will be some value $\lambda$. Consider strip i in the $j^{\text{th}}$ plane of the detector and events A and B, where the detector's response in event A was $n_{a}$ photoelectrons and the response in event B was $n_{b}$ photoelectrons. The likelihood $L$, of a data event corresponding to the same physical shower topology as a simulated library event can
therefore be calculated as

\begin{equation}
\log L = \sum\limits_{i=1}^{N_{\text{strips}}}  \log \left[ \int_{0}^{\infty} P\left( n^{i}_{\text{data}} | \lambda \right) P \left( n^{i}_{\text{lib}} | \lambda\right) d\lambda \right].
\label{eqn:NueLemLike}
\end{equation}

\noindent An example of a good and bad match from LEM can be seen in figure \ref{fig:nuematchevents}. 

\begin{figure}[!ht]
\centering
\includegraphics[height=0.60\textwidth]{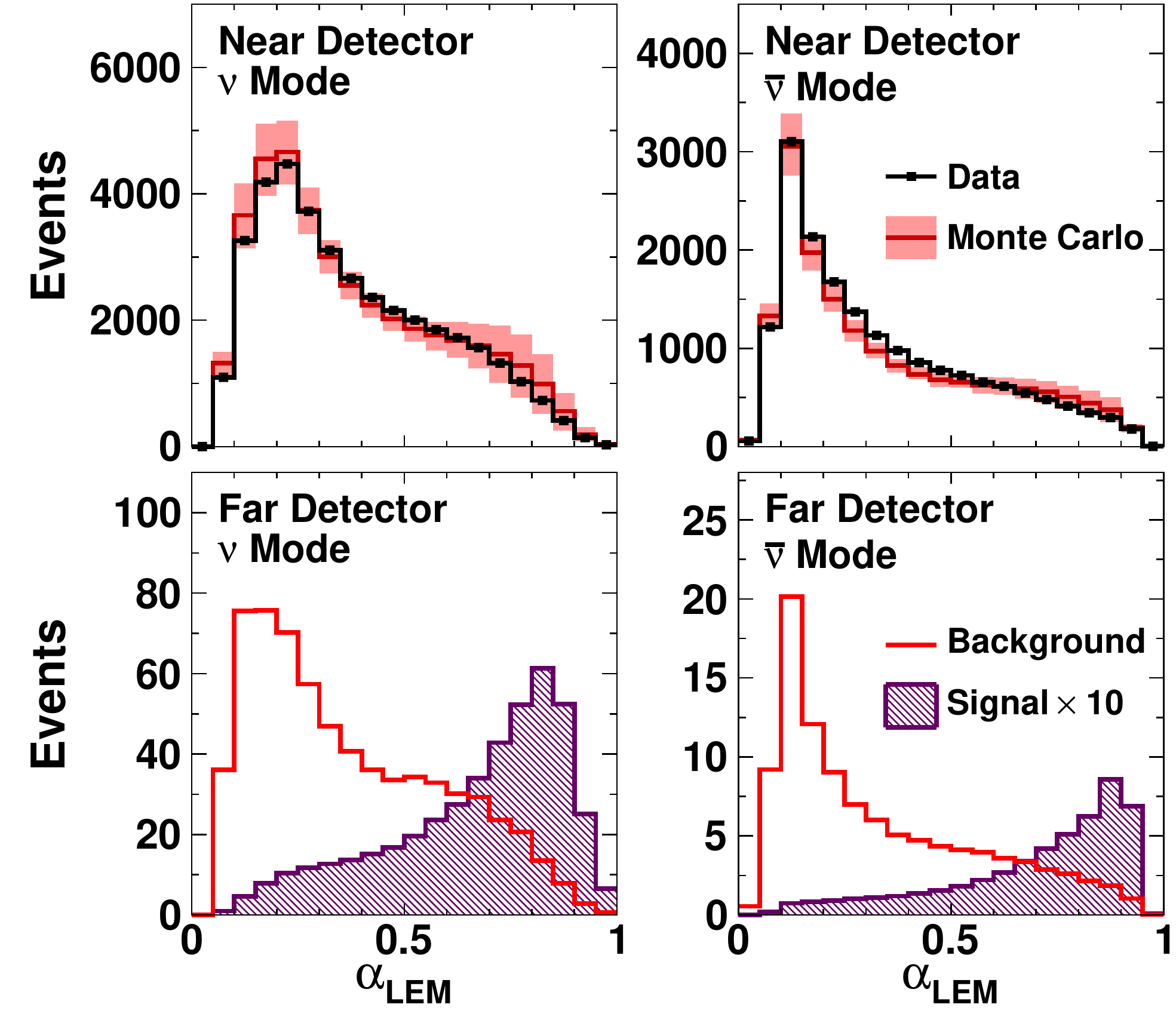}
\caption{Distributions of $\alpha_{\text{LEM}}$. The plots in the left column correspond to the neutrino dominated beam mode. The plots in the right column correspond to the anti-neutrino enhanced beam mode. The top row shows the distributions for ND selected events with a band about the simulation representing the systematic uncertainty. The bottom row shows the distributions for the predicted FD background and signal multiplied by 10}
\label{fig:lem}
\end{figure}

Three variables are constructed from the 50 best simulated events (signal or background), these are: the fraction of the events that are true \nue CC events, the average inelasticity (this is the amount of energy that goes into the hadronic shower) of the true \nue CC events, and the average fraction of charge that overlaps between the data event and each \nue CC library event. These three variables along with the reconstructed energy of the data event, are fed into a neural network which calculates a classification of how signal-like the data event is. A single variable, $\alpha_{\text{LEM}}$, is formed to quantify this, as shown in figure~\ref{fig:lem}. it's an output of an artificial neural network with several variables coming from the event comparisons. Events with $\alpha_{\text{LEM}} > 0.6$ are selected for analysis, this number was optimised to maximise the sensitivity to \nue appearance~\cite{ref:MINOSNuE2012}.
Candidate \nue-CC and \nuebar-CC events are required to fall within a fiducial volume and to be coincident in time ($50~\mu s$) and direction with the NuMI beam. Events are required to have shower-like topologies by rejecting events with tracks that are longer than 25 planes or extend more than 15 planes from a shower edge. 

With the absence of a \nue-CC and \nuebar-CC signal in the ND, the signal-selection efficiency can not be extrapolated from the ND events in the same way as the background estimate. By using real data, well-identified \numu-CC events are selected. By removing the energy deposited by the muon track~\cite{ref:AnnaThesis} these events can be used to calculated the signal selection efficiency by inserting the energy deposition from an electron with identical momentum to that of the removed muon. This allows one to effectively convert a well identified sample of \numu-CC and \numubar-CC data events into a sample of \nue-CC and \nuebar-CC data events. Using this method, the \nue-CC identification efficiency is found to be $(57.4\pm2.8)\%$ in the neutrino-dominated beam, and $(63.3\pm3.1)\%$ in the antineutrino-enhanced beam.

\subsection{Selection of neutral-current interactions}
The signal for a NC event is a diffuse hadronic shower, however \numu-CC interactions can also have large hadronic showers, if the inelasticity of the event is high ( most of the energy is given to the shower). The muon track may then be difficult to detect if it does not significantly extend beyond the hadronic shower. To achieve a high purity of NC events a number of selection cuts are used~\cite{ref:GemmaThesis}. An event is classified as a NC event if it has no tracks or if a track does not extend more than six planes past the end of the shower. The NC identification efficiency is 89\%, with 61\% purity, this is determined by taking an average over the energy spectrum for simulated events. However, this selection will identify 97\% of \nue CC interactions as NC events, which needs to be taken into account when searching for NC disappearance at the FD.

\subsection{Selection of atmospheric neutrinos}

\begin{figure}[!ht]
\centering
\includegraphics[height=0.40\textwidth]{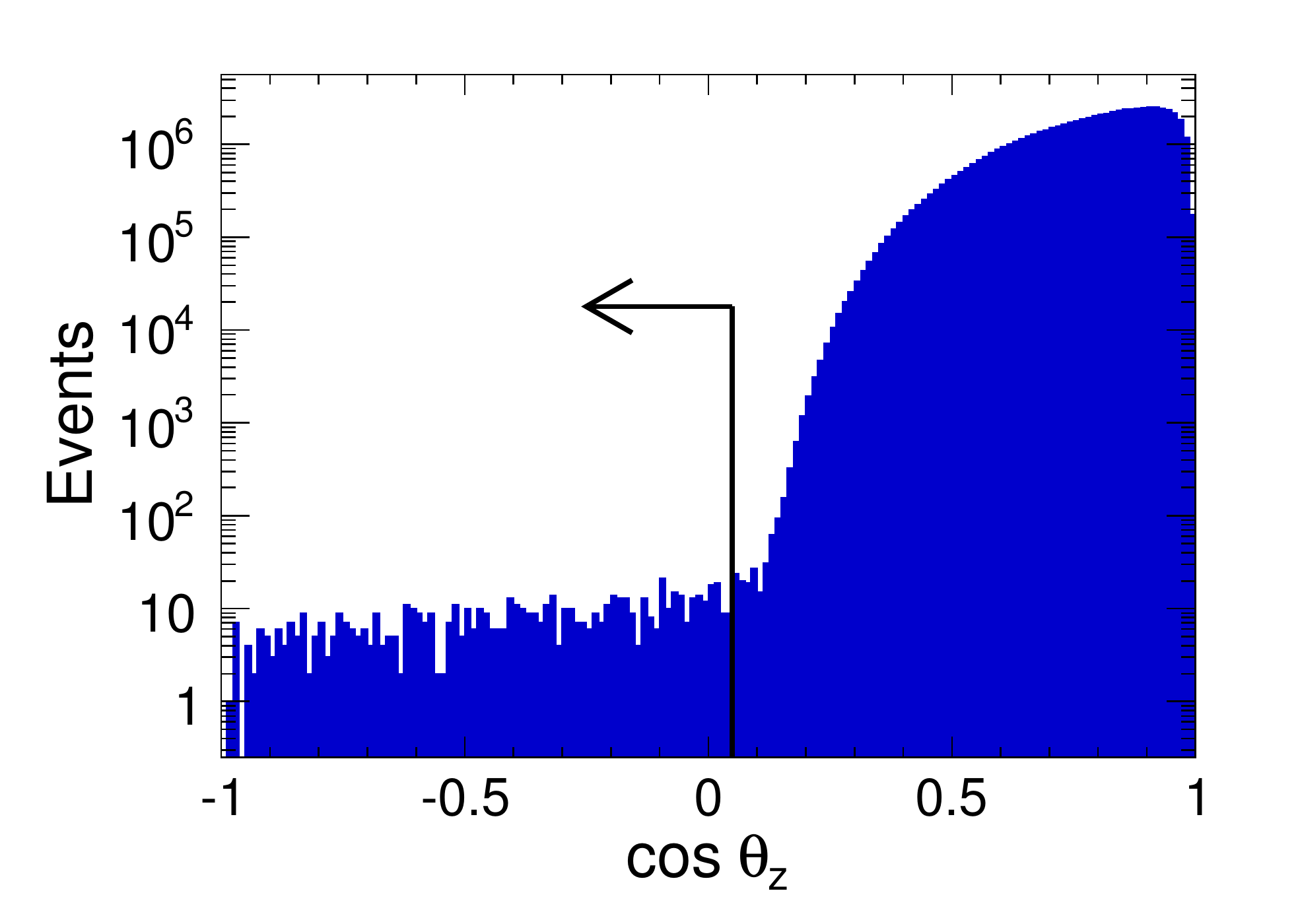}
\caption{Distribution of reconstructed zenith angle for muons with good timings and topology. In the range $\cos \theta_{z} > 0.10$ the observed rate of muons is dominated by the cosmic-ray background and falls steeply as the mean rock overburden increases. To minimise the background from cosmic-ray muons, events for analysis are required to satisfy $\cos \theta_{z} < 0.05$.}
\label{fig:FDatmoscut}
\end{figure}

Atmospheric neutrinos are selected as \numu-CC events in the MINOS FD outside of the 10~$\mu s$ window period when the NuMI beam is producing neutrinos~\cite{ref:MINOSAtmosPRD}. The atmospheric neutrino signal is separated from the cosmic-ray background using two characteristic signatures of atmospheric neutrino interactions: either a reconstructed vertex inside the fiducial volume; or a reconstructed upward-going or horizontal muon trajectory. 

The FD timing resolution on a hit by hit basis is 2.5~$ns$, which is enough to calculate the direction of a muon track inside the FD. For upward and horizontal angles, where the rock overburden
exceeds 14,000~m water-equivalent, the absorption of cosmic-ray muons by the earth is sufficiently high that the observed flux of muons is dominated by atmospheric muon neutrino interactions~\cite{ref:atmosOverburden}. At the Soudan mine (the location of the FD), upward-going tracks with a zenith angle $\cos\theta_{z} < 0.14$ are defined as upward-going and horizontal~\cite{ref:atmosSoudan}. Therefore, upward-going and horizontal provide a signature for atmospheric neutrinos. To further reduce the background the analysis requires  $\cos \theta_{z} < 0.05$ as shown in figure~\ref{fig:FDatmoscut}. 

For tracks where the end point lies inside the fiducial volume, the muon momentum is reconstructed from the measured track length; for exiting tracks, the momentum is obtained from the fitted track curvature. In both cases, the fitted curvature is used to determine the muon charge sign.

\section{Muon neutrino and antineutrino disappearance}

\begin{figure}[!ht]
\centering
\includegraphics[height=0.42\textwidth]{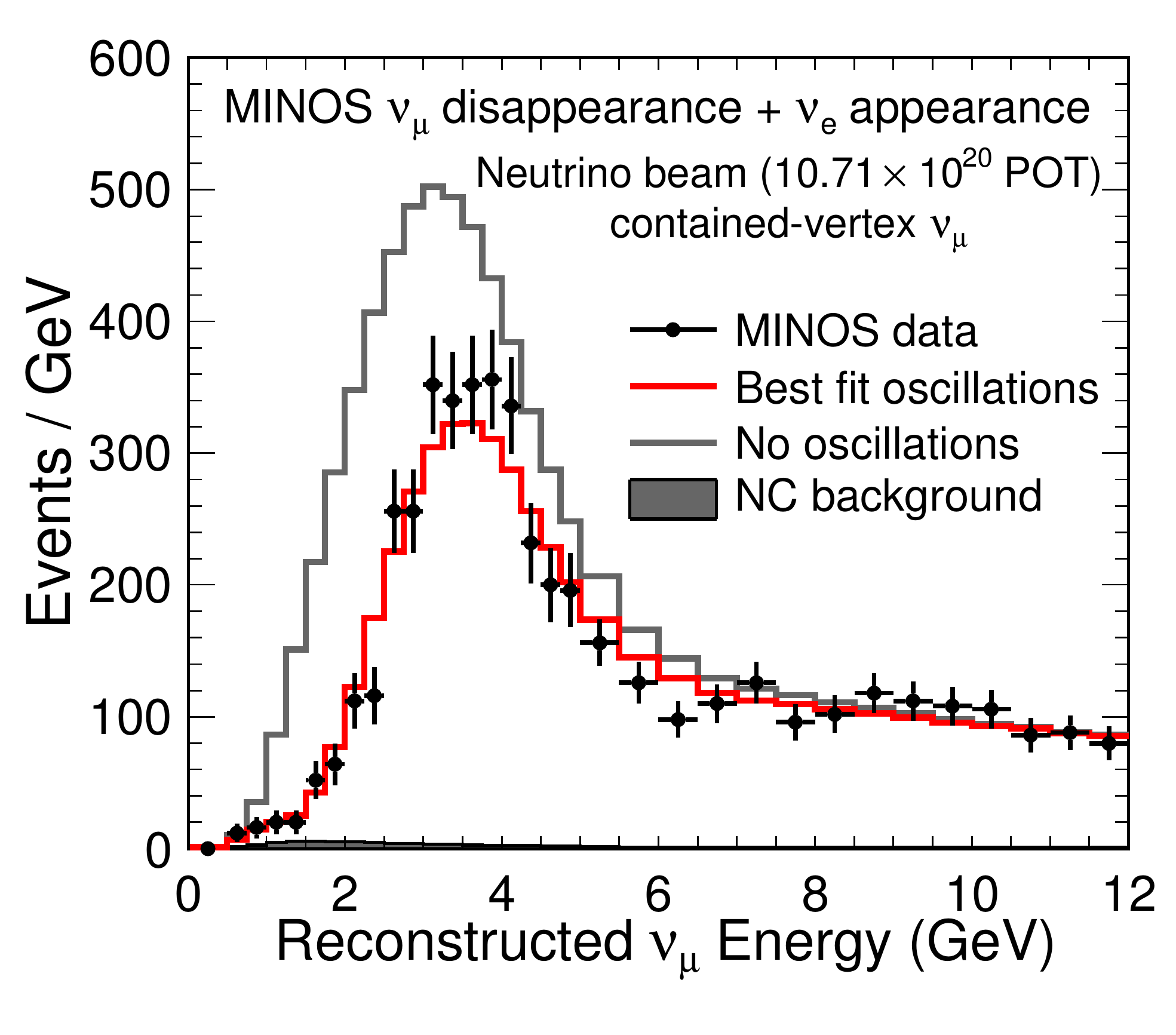}
\includegraphics[height=0.42\textwidth]{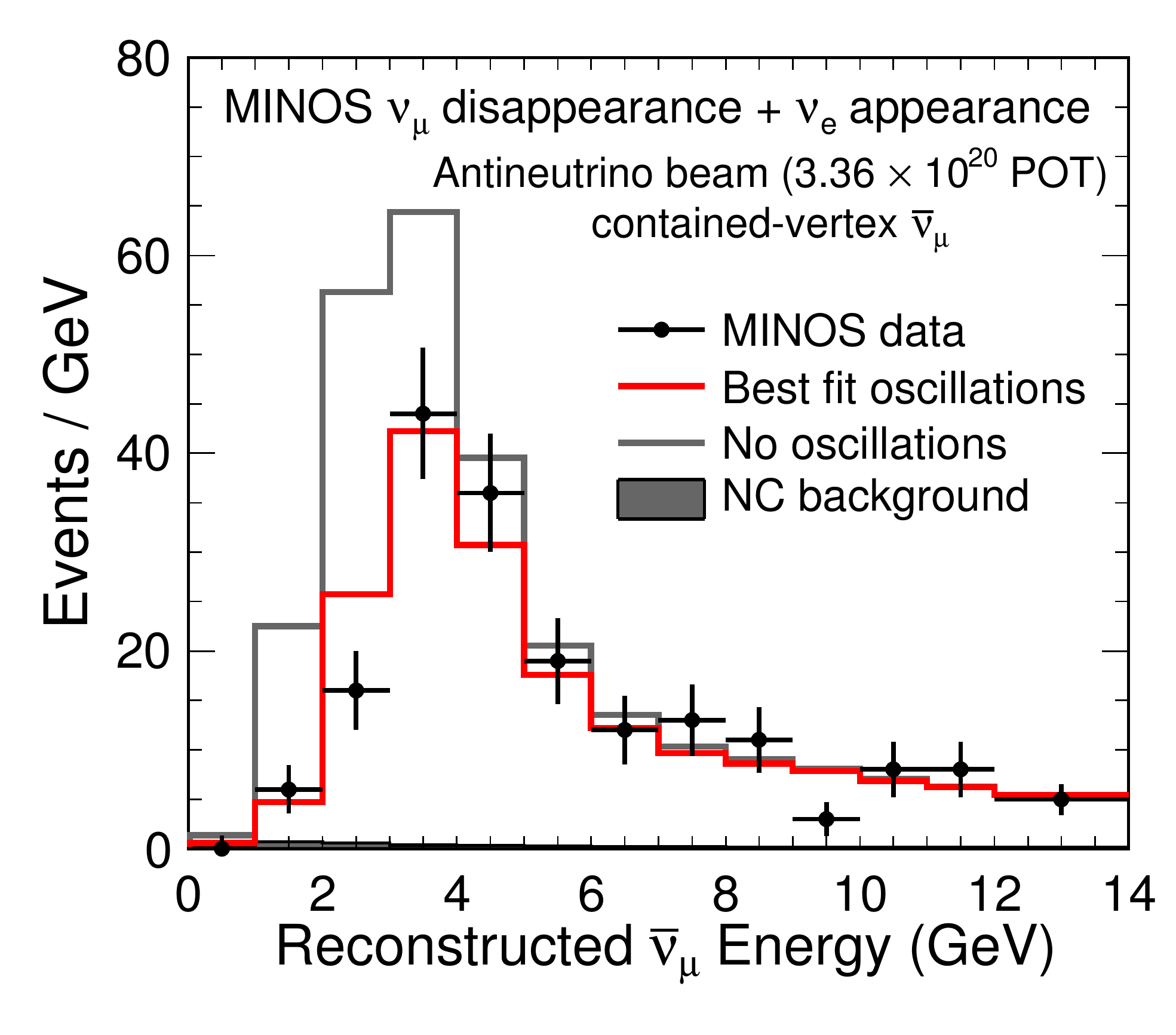}
\caption{The FD reconstructed neutrino energy spectrum for \numu (left) and \numubar (right). The light grey light shows the predicted FD energy spectrum if neutrinos did not oscillate. The red line indicates the best fit to the data. These energy spectra were fit along with with the atmospheric and \nue samples discussed in the later subsections. For details on the fit and the result of the neutrino oscillation parameters obtained see section~\ref{comboAnal}.} 
\label{fig:numu}
\end{figure}

MINOS can measure the atmospheric neutrino oscillation parameters  $\Delta m^{2}_{32}$ and $\theta_{23}$ by fitting the energy dependance of  \numu-CC and \numubar-CC disappearance. The FD sees the neutrino beam as a point source whereas the ND subtends a relatively large angle to the beam. Thus once a neutrino parent decays the ND would see a large spread in energies compared to the FD which preferentially selects higher energy neutrinos at a smaller angle from the direction of the parent hadron. MINOS pioneered the technique of resolving this kinematic issue by constructing a beam matrix that allows one to convert an energy spectrum observed in the ND to the corresponding spectrum observed at the FD~\cite{ref:JustinThesis}. By using the ND data as a constraint many systematics which affect both detectors cancel to first order. Only systematics affecting both detectors differently become a significant source of uncertainty; primarily reconstruction efficiencies and miscalibrations of the neutrino energy measurement in the detectors ~\cite{ref:StephenThesis}.  Figure~\ref{fig:numu} shows the reconstructed energy spectrum at the FD for \numu and \numubar neutrino events compared to two predictions, if  there were no oscillations and a best fit to the FD data. 

There is an uncertainty on the relative normalisation of the selected Near and Far Detector event samples which is dominated by differences in the reconstruction and selection efficiencies between the two detectors, as well as relative uncertainties on fiducial mass and live time; this uncertainty is found to be 1.6\%. There are two different uncertainties on the measurement of hadronic shower energy~\cite{ref:BackhouseThesis}, these are the relative mis-modelling of the energy scale between the two detectors as well as the absolute mis-modelling. It was found that the relative uncertainty for the ND is 1.9\% and the FD is 1.1\%. The absolute mis-modelling comes from the uncertainties on the modelling of hadronic showers which is fully correlated bin to bin in reconstructed energy and has an energy dependence of the form $\sigma_{\text{shw}} = 6.6\% + (3.5\%) \times \text{exp}\left( -E_{\text{reco}} /1.44 \text{GeV} \right )$~\cite{ref:shower}.

\begin{figure}[!ht]
\centering
\includegraphics[height=0.55\textwidth]{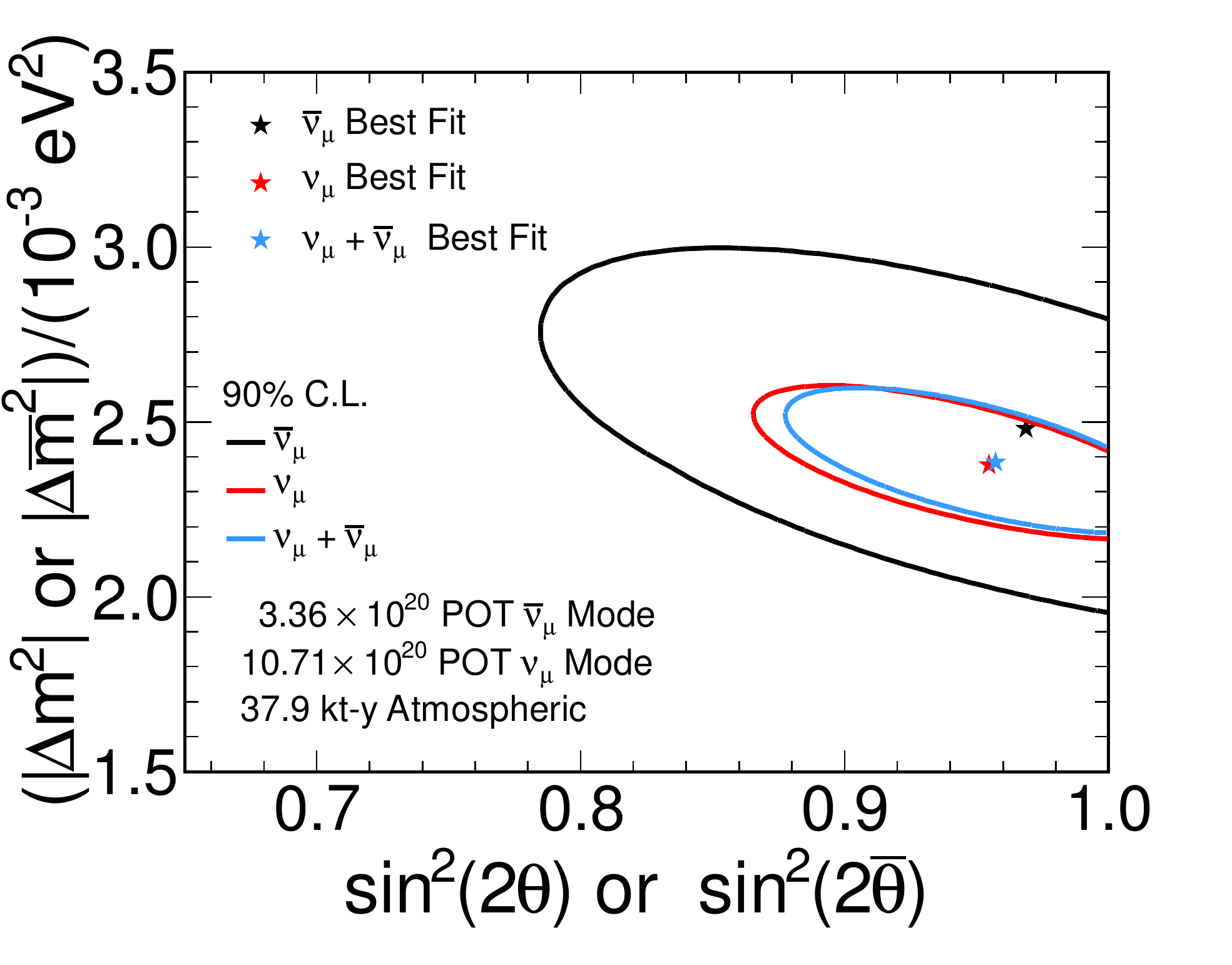}
\caption{The allowed region for antineutrino oscillation parameters (black line), compared to the region measured with neutrinos alone (red line) and the region measured using both neutrinos and antineutrinos under the assumption they have the same parameters (blue line).}
\label{fig:numubar}
\end{figure}

A previous two-flavour analysis~\cite{ref:MINOS2flav2013} of \numu and \numubar disappearance using the combined accelerator and atmospheric data from MINOS yielded $|\Delta m^{2}| = 2.41^{+0.09}_{-0.10} \times 10^{-3} \, \text{eV}^{2}$ and $\sin^{2} 2\theta = 0.950^{+0.035}_{-0.036}$. A symmetry in CPT requires that neutrinos and anti-neutrinos oscillate in an identical way, thus their oscillation parameters should be identical. With the ability to distinguish the lepton number of a neutrino MINOS can measure the oscillation parameters for anti-neutrinos and neutrinos separately. Using both atmospheric and beam anti-neutrinos MINOS measures the oscillation parameters to be $|\Delta \overline{m}^{2}| = 2.50^{+0.23}_{-0.35} \times 10^{-3} \, \text{eV}^{2}$ and $\sin^{2} 2\overline{\theta} = 0.97^{+0.03}_{-0.08}$~\cite{ref:MINOS2flav2013} which is in good agreement with the parameters measured from \numu oscillations as shown in figure~\ref{fig:numubar}.

\subsection{Atmospheric \numu and \numubar disappearance}

\begin{figure}[!ht]
\centering
\includegraphics[height=0.55\textwidth]{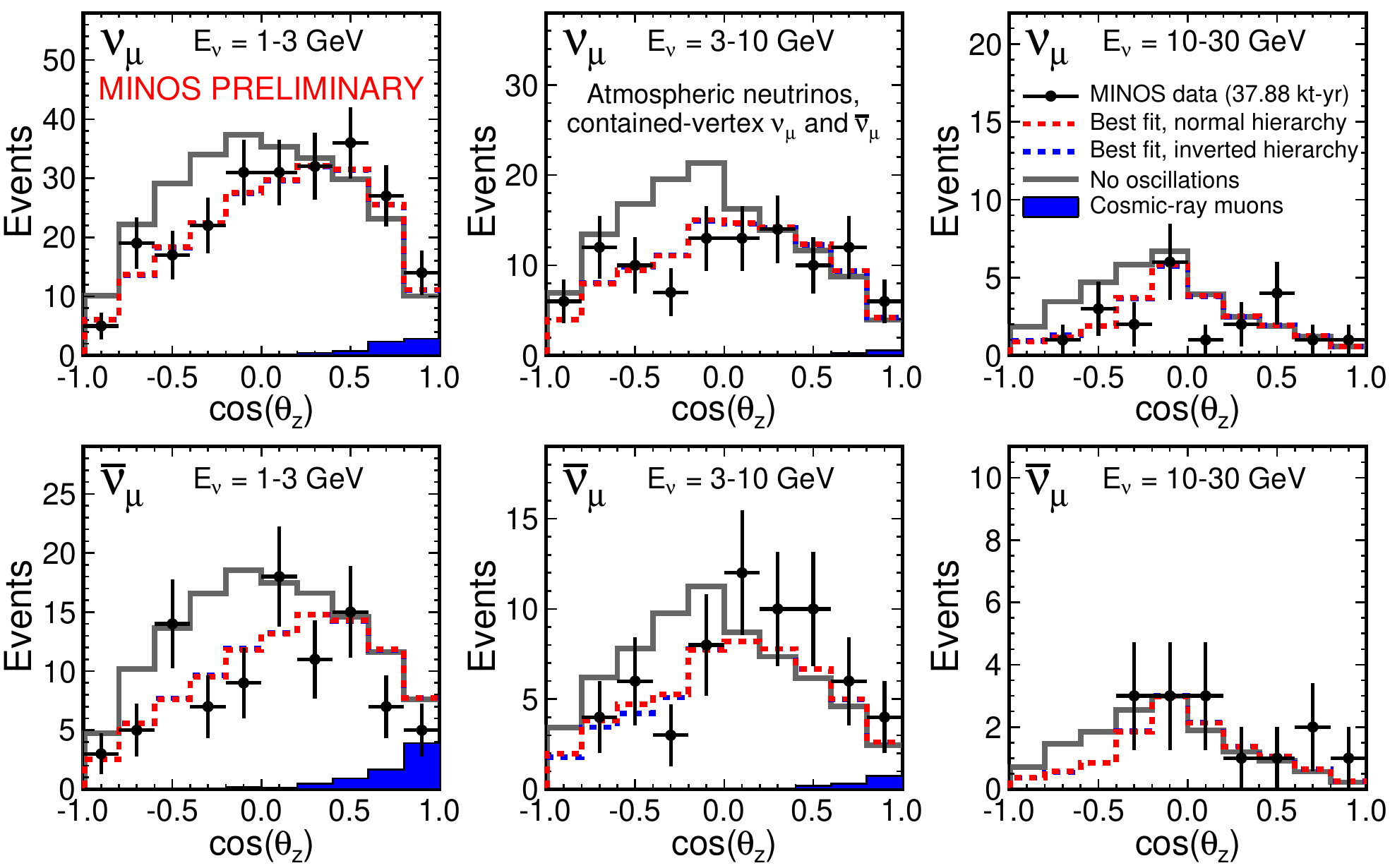}
\caption{The distribution of reconstructed atmospheric \numu and \numubar CC interactions over various energy ranges observed at the FD, compared to the expectation with and without neutrino oscillation. Two different fits are performed for the normal and inverted hierarchy of $\Delta m^{2}_{32}$. The energy of the neutrino event is the sum of muon momentum (from range and/or curvature) and total shower energy.}
\label{fig:atmosfit}
\end{figure}

MINOS is the first experiment to probe the resonance predicted to occur in multi-GeV, upward-going atmospheric neutrinos which travel through the earth's mantle for both neutrino and anti-neutrino events on an event by event basis. The atmospheric events are separated into samples of contained-vertex and non-fiducial muons for neutrinos and anti-neutrinos, figure~\ref{fig:atmosfit} shows the atmospheric samples containing events with a contained-vertex. By measuring the \numu-CC and \numubar-CC interactions separately this allows MINOS to gain sensitivity to the mass hierarchy and $\theta_{23}$ octant. The difference between the inverted and normal mass hierarchy in figure~\ref{fig:atmosfit} is very marginal.

The neutrino events are binned as a function of $\log_{10} \left( E \right)$ and $\cos \theta_{z}$, where E is the reconstructed energy of the event in GeV and $\theta_{z}$ is the zenith angle of the muon track, this binning gives enhanced sensitivity to the MSW resonance. A sample of contained-vertex showers are also selected from the data, composed mainly of NC, \nue-CC and \nuebar-CC interactions. They are used to constrain the overall flux normalization. For atmospheric neutrinos, the earth is modelled by four layers of constant electron density using the PREM model~\cite{ref:PREM}. Comparisons to a more detailed 52 layer model yielded very similar results and so the extra computational time was avoided by using the simple four layer model.

\section{\nue and \nuebar appearance}

By searching for \nue and \nuebar appearance at the FD MINOS can perform a measurement of $\theta_{13}$. The POT exposure for this data set is  $10.6 \times 10^{20}$ protons-on-target (POT) using a \numu-dominated beam and $3.3 \times 10^{20} $ POT using a \numubar-enhanced beam. Neutrino events with $\alpha_{\text{LEM}} > 0.6$ are selected for analysis in the \numu-domainted beam mode and in the \numu-enhanced beam. Neutrino events with $\alpha_{\text{LEM}} < 0.6$ are considered background-like and therefore insensitive to \nue and \nuebar appearance. The background consists of three components: NC interactions, CC-\numu and \numubar interactions, and the intrinsic \nue component in the beam. The relative contribution between the ND and FD is different for all of these components, since they are affected differently by oscillation, and the kinematics of the production in the beam are different. Each background must be individually measured. By changing the configuration of the NuMI beam (low, medium or pseudo-high) one can measure these backgrounds. The relative contributions of the background components changes in a well understood way~\cite{ref:JoaoThesis}. Neutrino events with $\alpha_{\text{LEM}} < 0.6$ are used to provide validation to the analysis procedure; MINOS uses ND neutrino events with $\alpha_{\text{LEM}} < 0.5$ to predict FD event yields. Comparing these yields to that found in the FD agree to within $0.3\sigma$ ($0.6\sigma$) of the statistical uncertainty for the data sample in the \numu (\numubar) beam modes.  

\begin{figure}[!ht]
\centering
\includegraphics[height=0.80\textwidth]{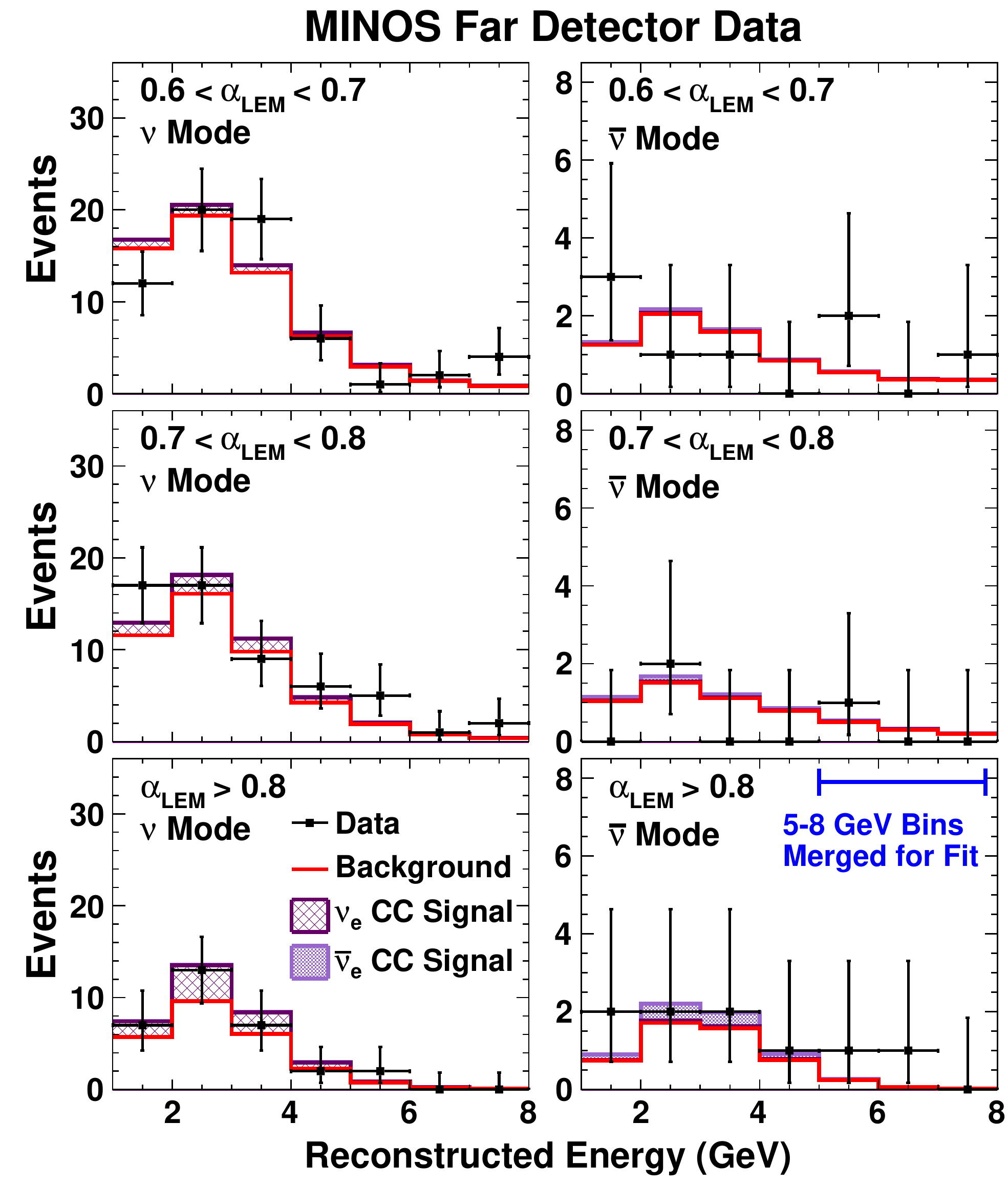}
\caption{The reconstructed neutrino energy distributions for three $ \alpha_{\text{LEM}}$ ranges. The events with energy greater than 5 GeV are combined into a single bin for the fits. The vertical bars through the data points denote statistical uncertainties. The signal prediction assumes $\sin^{2}(2\theta_{13}) = 0.051$, $\Delta m^{2}_{32} > 0$ (normal hierarchy), $\delta_{CP} = 0$ and $\theta_{23} = \pi/4$ (maximal mixing). The left corresponds to data taken in \numu-dominated beam mode samples and the right is from \numubar-enhanced beam mode samples.}
\label{fig:nuefit}
\end{figure}

For an appearance analysis one has to consider a three-flavour neutrino oscillation probability that includes matter effects. The fit is done simultaneously for data from both \numu-dominated beam mode and \numubar-enhanced beam mode samples. A total of 127.7 background events are expected at the FD in the neutrino-dominated beam, and 17.5 events in the
antineutrino-enhanced beam. In the data, 152 and 20 events are observed, respectively and their energy distributions can be seen in figure~\ref{fig:nuefit}.

MINOS cannot distinguish \nue and \nuebar events directly, however, the relative number of neutrino and antineutrino interactions in the neutrino-dominated and anti-neutrino-enhanced beams is well known. The parameter measured is $2\sin^{2}(2\theta_{13})\sin^{2}(\theta_{23})$ while the values $\Delta m^{2}_{32}$ and $\delta_{CP}$ remain fixed. The fit is ran over both hierarchies and all possible values of $\delta_{CP}$. For more details on this analysis see reference~\cite{ref:MINOSNuE2012} . The fit is performed using the 15 bins formed by three bins of $\alpha_{\text{LEM}}$ and five bins of energy as seen in figure~\ref{fig:nuefit}, note final three bins in energy from 5-8~GeV were merged for the fit.

\begin{figure}[!ht]
\centering
\includegraphics[height=0.80\textwidth]{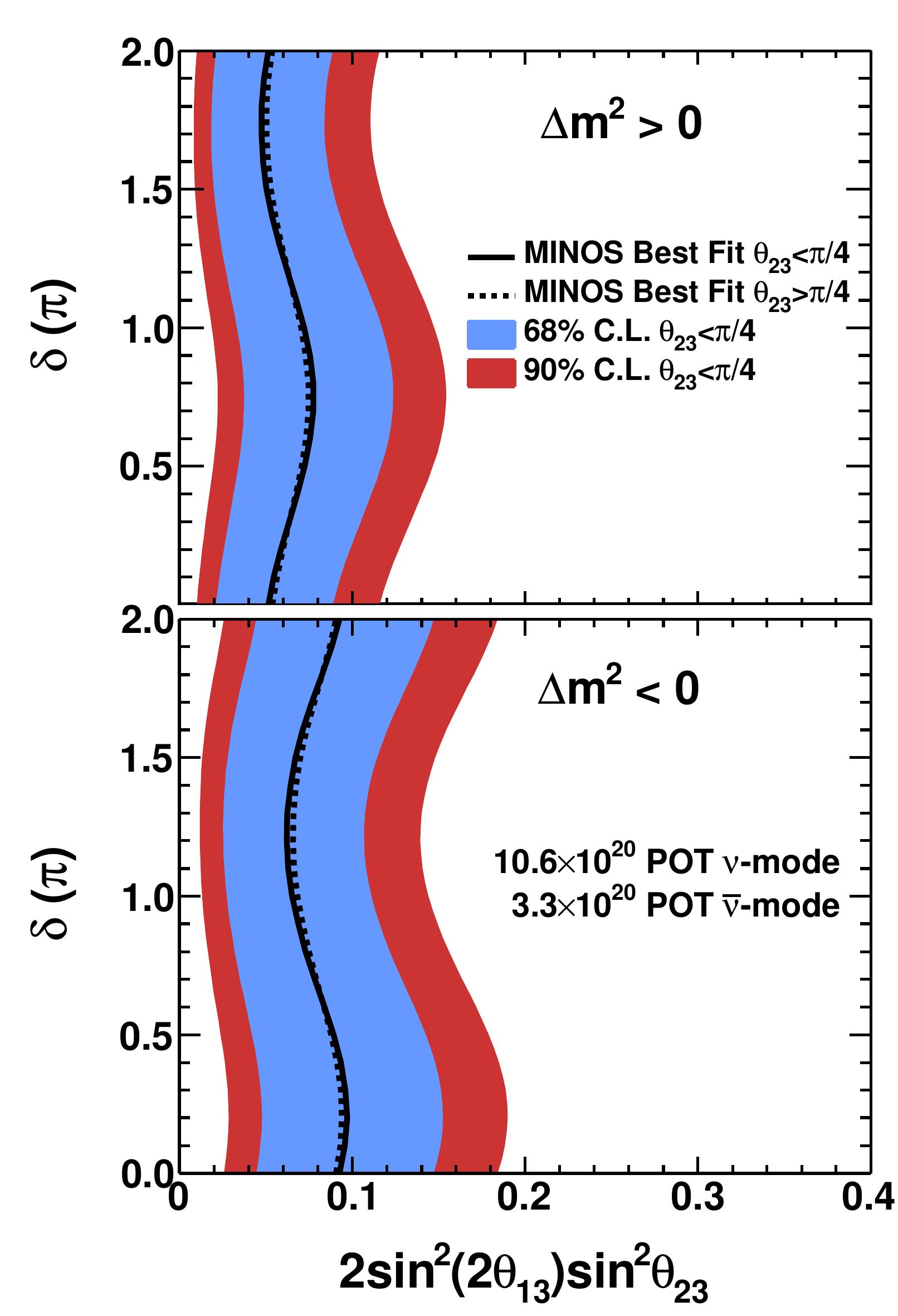}
\caption{The allowed regions for $2\sin^{2}(2\theta_{13})\sin^{2}(\theta_{23})$.}
\label{fig:nue}
\end{figure}

MINOS finds that the data allow for a value of $2\sin^{2}(2\theta_{13})\sin^{2}(\theta_{23}) = 0.051^{+0.038}_{-0.030}$ for the normal hierarchy with $\delta_{CP} = 0$ and $2\sin^{2}(2\theta_{13})\sin^{2}(\theta_{23}) = 0.093^{+0.054}_{-0.049}$ for the inverted hierarchy with $\delta_{CP} = 0$, in both cases $\theta_{23} < \pi/4$, this can be seen in figure~\ref{fig:nue}. This the first ever \nuebar appearance search in a long-baseline \numubar beam.

\section{A Combined Three Flavour Analysis}
\label{comboAnal}

To gain maximum sensitivity to the mass hierarchy and $\delta_{CP}$, one needs to perfrom a full three flavour analysis combining \numu and \numubar disappearance with \nue and \nuebar appearance. For this one needs to account for both disappearance and appearance of neutrinos oscillations at both detectors. 

\begin{figure}[!ht]
\centering
\includegraphics[height=0.6\textwidth]{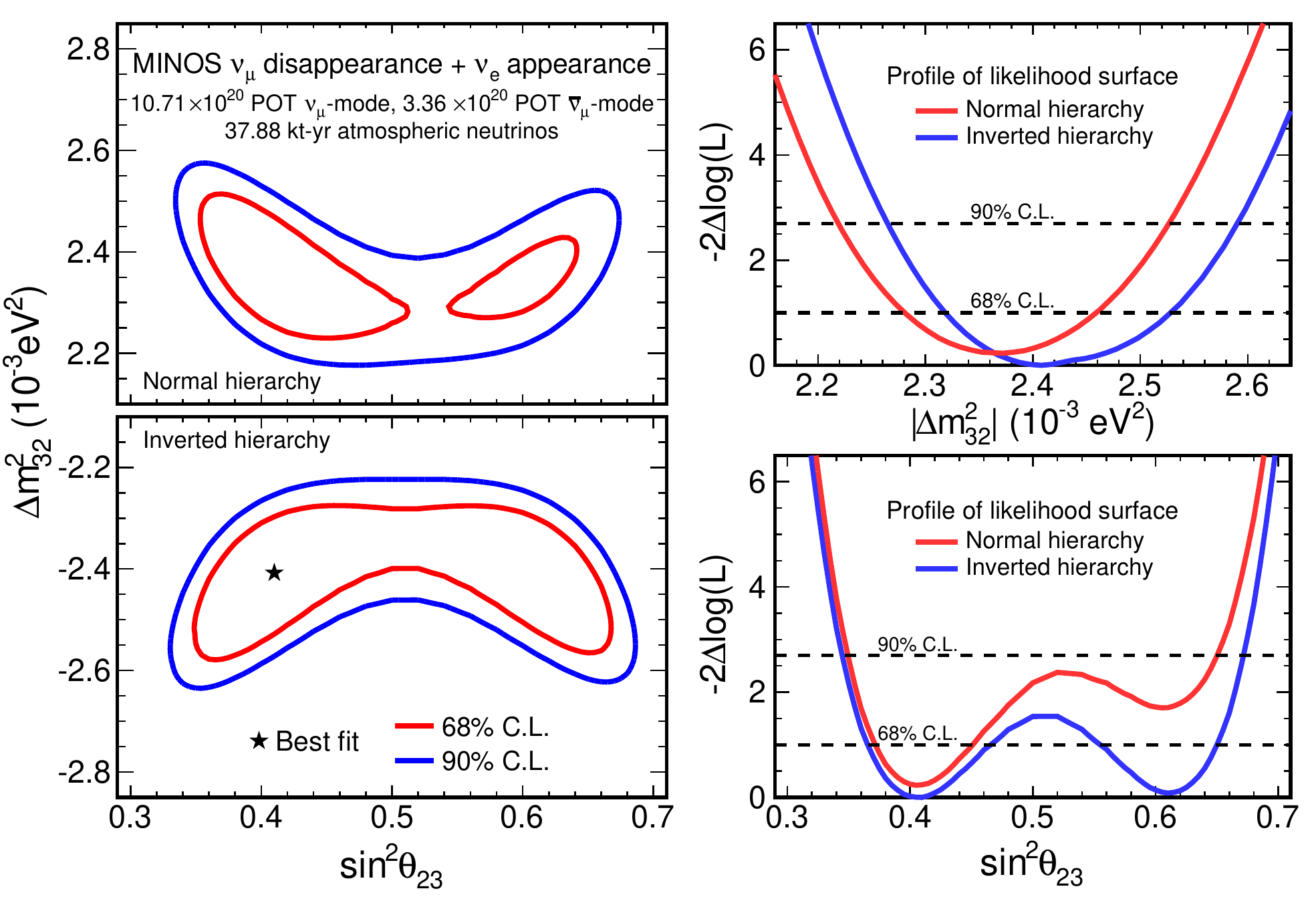}
\caption{The left panels show the 68\% and 90\% confidence limits on $\Delta m^{2}_{32}$ and $\sin^{2} \theta_{23}$ for the normal hierarchy (top) and inverted hierarchy (bottom). At each point in this parameter space, the likelihood function is maximised with respect to $\sin^{2} \theta_{13}$, $\delta_{cp}$ and the systematic parameters constraints in the fit. The likelihood surface is calculated relative to the overall best fit, which is indicated by the star. The right panels show the 1D likelihood profiles as a function of $\Delta m^{2}_{32}$ and $\sin^{2} \theta_{23}$ 
for each hierarchy. The horizontal dotted lines indicate the 68\% and 90\% confidence limits.}
\label{fig:combinedcontours}
\end{figure}

For a full three-flavour fit MINOS uses an accelerator neutrino data set comprising exposures of $10.71 \times 10^{20}$ protons-on-target (POT) using a \numu-dominated beam and $3.36 \times 10^{20} $ POT using a \numubar-enhanced beam. Both sets were acquired in the low energy NuMI beam configuration. MINOS also collected $37.88$ kt-years of atmosphere neutrino data. 

The oscillation parameters are determined by applying a maximum likelihood fit to the data. The parameters $\Delta m^{2}_{32}$, $\sin^{2} \theta_{23}$, $\sin^{2} \theta_{13}$ and $\delta_{CP}$ are varied in the fit with an external constraint on the mixing angle $\sin \theta_{13} = 0.0242 \pm 0.0025$ calculated from a weighted average of the latest published results from the reactor experiments Daya Bay~\cite{ref:DayaBay}, RENO~\cite{ref:Reno} and Double Chooz~\cite{ref:DoubleChooz}. The constraints are included by adding a Gaussian prior penalty term onto the likelihood during the fit. The solar parameters are kept at the fixed values of $\Delta m^{2}_{21} = 7.54 \times 10^{-5} \, \text{eV}^{2}$ and $\sin^2\theta_{12} =  0.037$~\cite{ref:Fogli}. To test the impact of the solar parameters on the fit they were varied within their uncertainties and the effect was negligible on the final results.

\begin{figure}[!ht]
\centering
\includegraphics[height=0.60\textwidth]{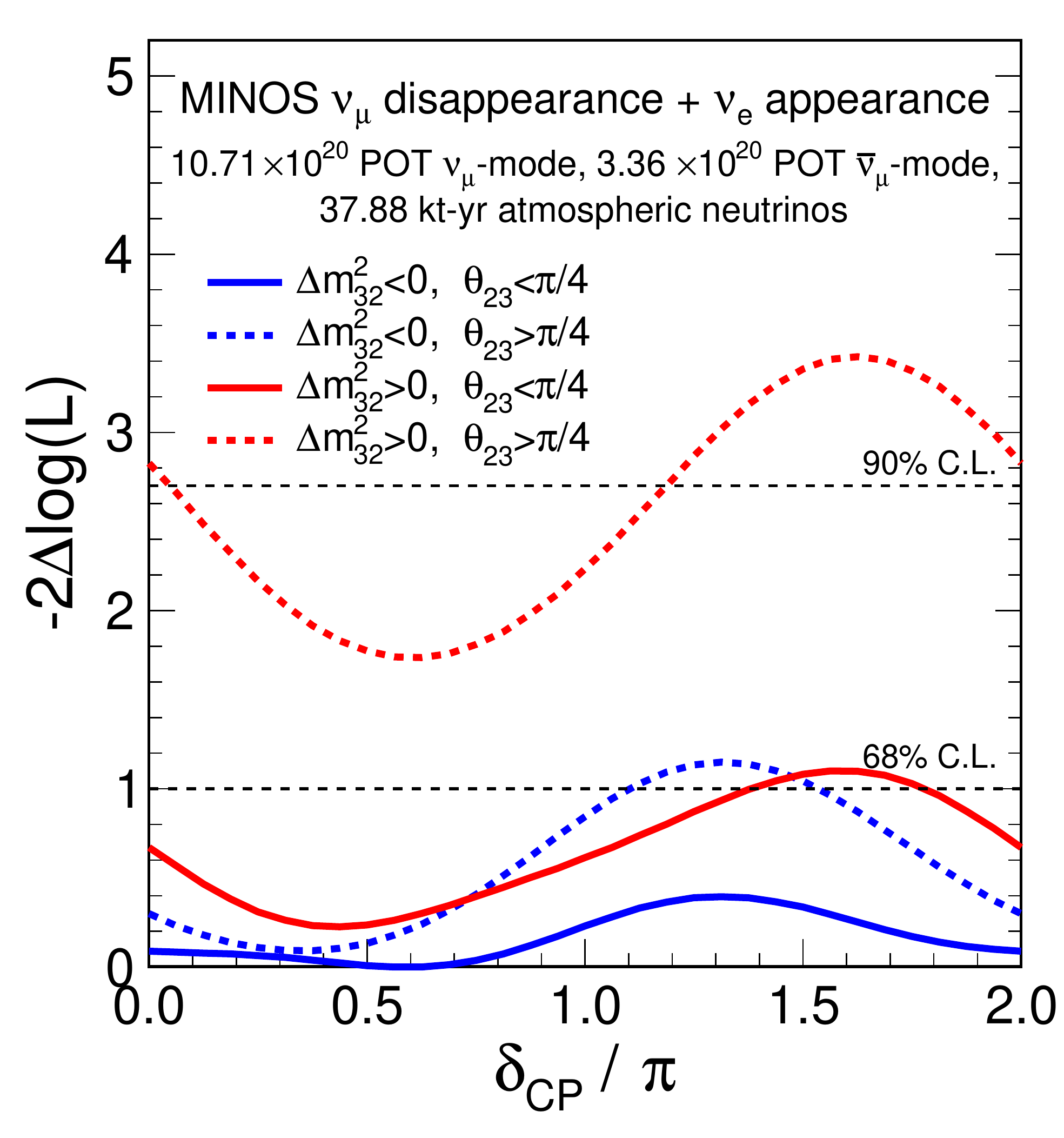}
\caption{The 1D likelihood profile as a function of $\delta_{CP}$ for each combination of mass hierarchy and $\theta_{23}$ octant. For each value of $\delta_{CP}$ the likelihood was maximised with respect to $\sin^{2} \theta_{13}$, $\sin^{2} \theta_{23}$ and $\Delta m^{2}_{32}$.}
\label{fig:deltafit}
\end{figure}

Figure~\ref{fig:combinedcontours} shows the 2D confidence limits on $\Delta m^{2}_{32}$, $\sin^{2} \theta_{23}$ as well as the 1D profiled confidence limits on $\Delta m^{2}_{32}$ and $\sin^{2} \theta_{23}$ separately. One can see the sensitivity to the octant of $\theta_{23}$ through the inclusion of the atmospheric neutrino sample.  The 68\% (90\%) confidence limits (C.L.) on these parameters are calculated by taking the range of negative log-likelihood values with $-2 \ln L$ < 1.00 (2.71) relative to the overall best fit. This yields $| \Delta m^{2}_{32} |= [2.28-2.46] \times 10^{-3} \, \text{eV}^{2}$ at 68\% C.L and $\sin \theta_{23} = 0.35-0.65$ at 90\% C.L in the normal hierarchy; $| \Delta m^{2}_{32} |= [2.32-2.53] \times 10^{-3} \, \text{eV}^{2}$ at 68\% C.L and $\sin \theta_{23} = 0.34-0.67$ at 90\% C.L in the inverted hierarchy. The data disfavour maximal mixing $\theta_{23} = \pi /4$ by  $-2 \ln L$ < 1.54. These results give the most precise measurement made on $\Delta m^{2}_{32}$ to date.

Figure \ref{fig:deltafit} shows the 1D C.L of the profiled likelihood surface for value of $\delta_{CP}$ for each of the four possible combinations (hierarchy and octant of $\theta_{23}$). The data disfavour 36\% (11\%) of the parameter space defined by $\delta_{CP}$, the $\theta_{23}$ octant, and the mass hierarchy at 68\% (90\%) C.L.

\section{NSI Interactions}
Non-standard interactions (NSI)~\cite{ref:NSI1,ref:NSI2,ref:NSI3} could occur between muon or tau neutrinos and matter, and could alter the flavour composition of a neutrino beam as it propagates through the earth's crust. Searches for NSI have already been performed with atmospheric neutrinos~\cite{ref:NSIsuperK}. Using a two-flavour approach one can write the probability for muon neutrino survival as:

\begin{figure}[!ht]
\centering
\includegraphics[height=0.5\textwidth]{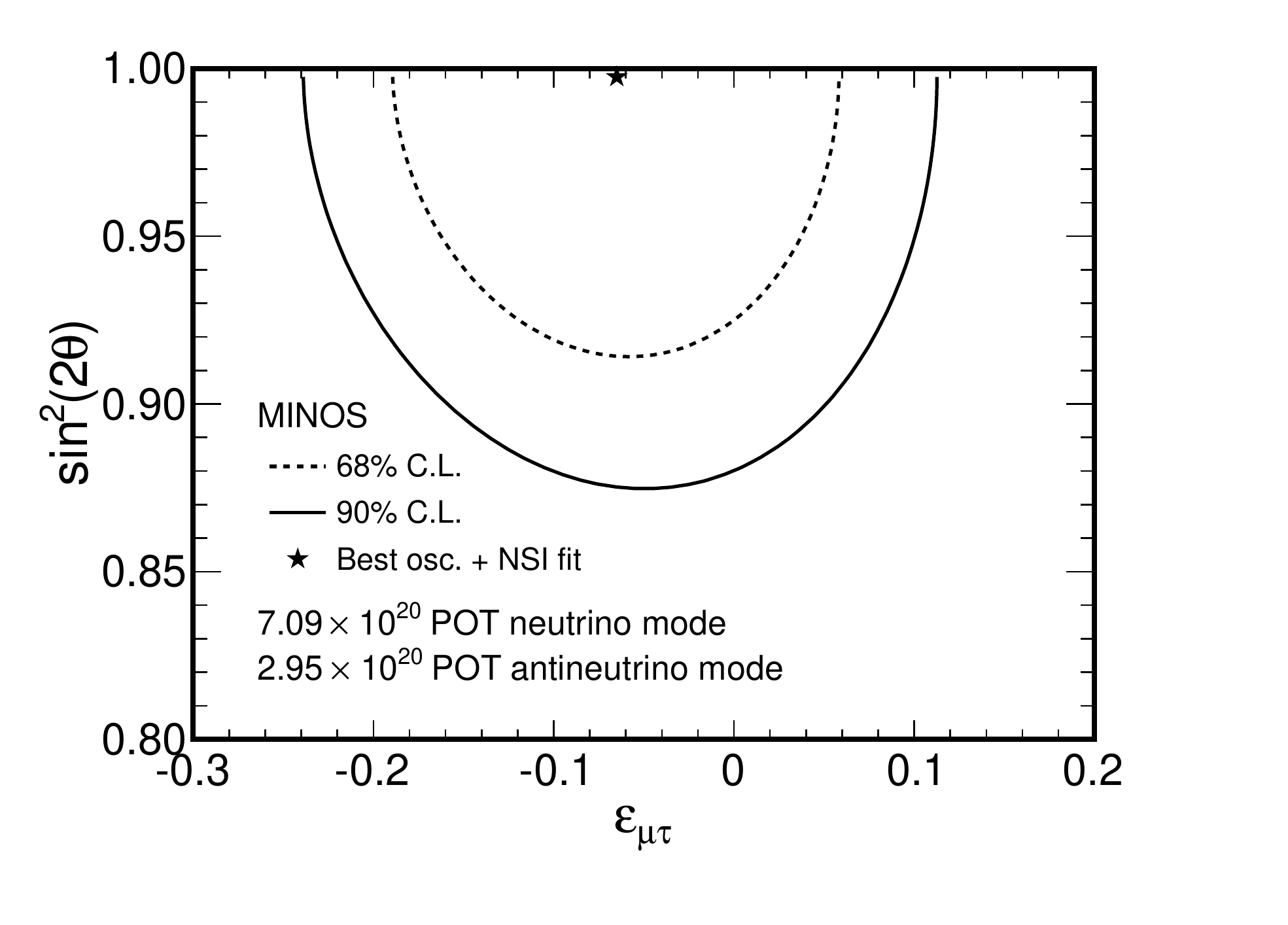}
\includegraphics[height=0.5\textwidth]{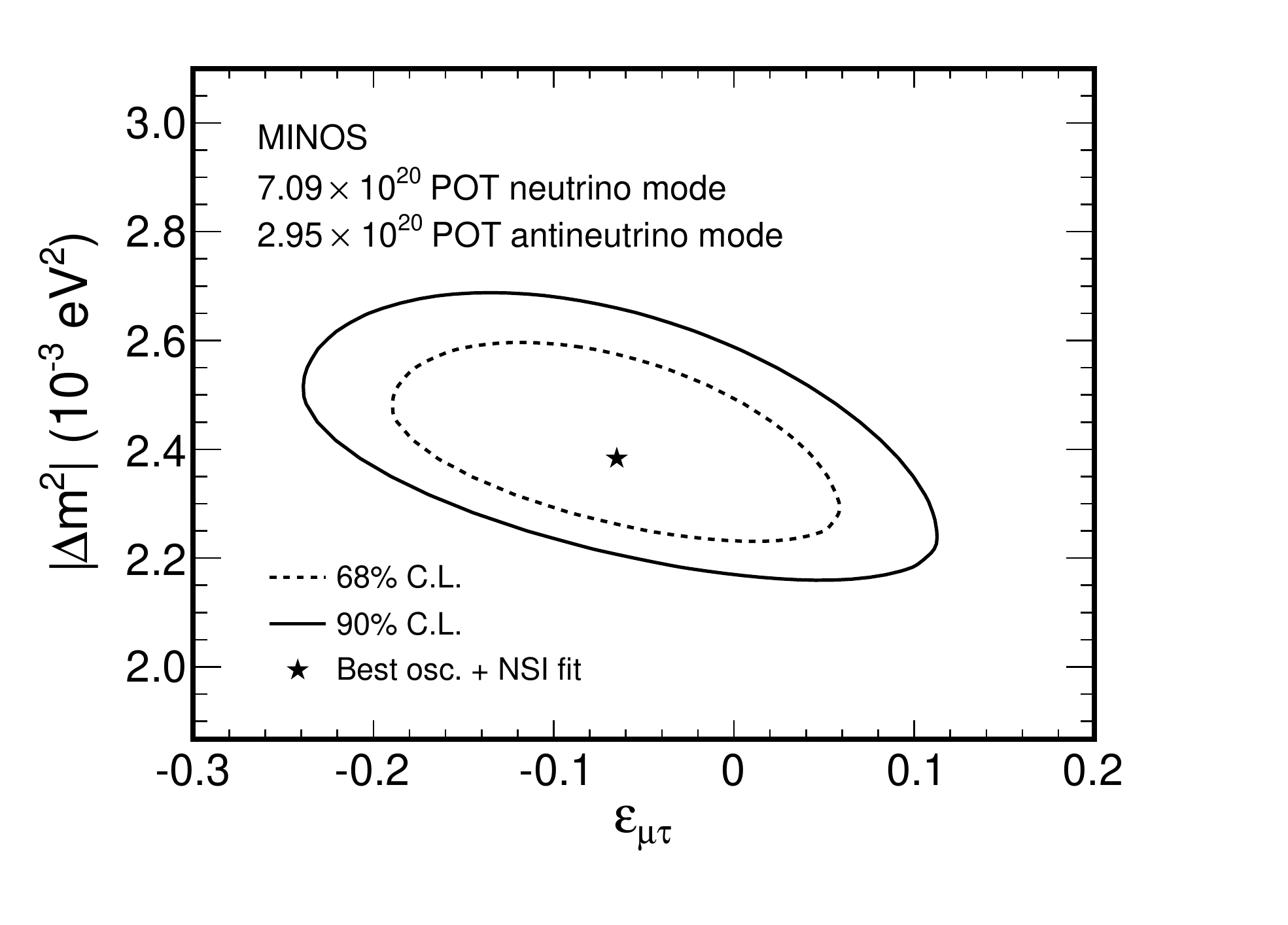}
\caption{2D contours for both 68\% and 90\%, showing the allowed regions of parameter space for $\epsilon_{\nu\tau}$ from the fit to the data through muon disappearance. }
\label{fig:NSIresults}
\end{figure}

\begin{align}
P \left( \nu_{\mu} \rightarrow \nu_{\mu} \right) = 1 - \left[ 1 - \cos^{2} \left( 2 \theta \right) \frac{L^{2}_{m}}{L^{2}_{0}} \right] \sin^{2} \left( \frac{L}{L_{m}}\right), 
\label{eqn:NSI1}
\end{align}

where $L$ is the neutrino path length and $L_{m}$ is defined as the NSI  matter oscillation length defined as:

\begin{align}
L_{m} \equiv \frac{L_{0}}{\left[ 1 \pm 2 \sin \left( 2 \theta \right) L_{0} \epsilon_{\mu \tau} |V| + \left( L_{0}\epsilon_{\mu \tau}|V|^{2}\right) \right]^{\frac{1}{2}}},
\label{eqn:NSI2}
\end{align}

\noindent where $L_{0} \equiv 4E/\Delta m^{2}$.The $\pm$ signs in equation~\ref{eqn:NSI2} arise from the matter potential, $V$, which is positive for neutrinos and negative for antineutrinos. The parameter $\epsilon_{\mu \tau}$ is real-valued and carries its own sign. A positive value of $\epsilon_{\mu\tau}$ implies that the neutrino disappearance probability is greater than the antineutrino disappearance probability, and vice versa.

The results presented here are based on an exposure of $7.09 \times 10^{20}$ protons on target (POT) in neutrino mode, combined with a $1.7 \times 10^{20}$ POT exposure in antineutrino mode. Due to the opposite sign of the matter potential in equation~\ref{eqn:NSI2} for neutrinos and antineutrinos, NSI, if present, will alter the survival probability of neutrinos and antineutrinos in opposite directions. This analysis has identical event selection as for the \numu disappearance analysis, however the fit takes into account perturbations from the standard three-flavour formalism brought about by NSI. The best fit parameters from this procedure are found to be  $\Delta m^{2} = 2.39^{+0.14}_{-0.11} \times 10^{-3} \, \text{eV}^{2}$, $\sin^{2} 2\theta = 1.00^{+0.00}_{-0.06}$ and $\epsilon_{\mu\tau} = -0.07^{+0.08}_{-0.08}$ with the allowed region $-0.20 < \epsilon_{\mu\tau} < 0.07$ (90\% C.L.). The systematic uncertainties incorporated into the penalty terms when maximising the likelihood have a negligible effect on the fit. Figure~\ref{fig:NSIresults} shows the allowed regions of the fit parameters, these are 2D C.L contours, produced by profiling the likelihood with respect to the other parameters. 

This is the first direct search for nonstandard interactions with high-purity samples of both neutrinos and antineutrinos conducting a simultaneous fit to neutrino and antineutrino energy spectra of conventional $\numu \rightarrow \nu_{\tau}$ oscillations with an additional NSI matter effect. This result is consistent with the null hypothesis of no NSI.

\section{Sterile Neutrinos}

\begin{figure}[!ht]
\centering
\includegraphics[height=0.40\textwidth]{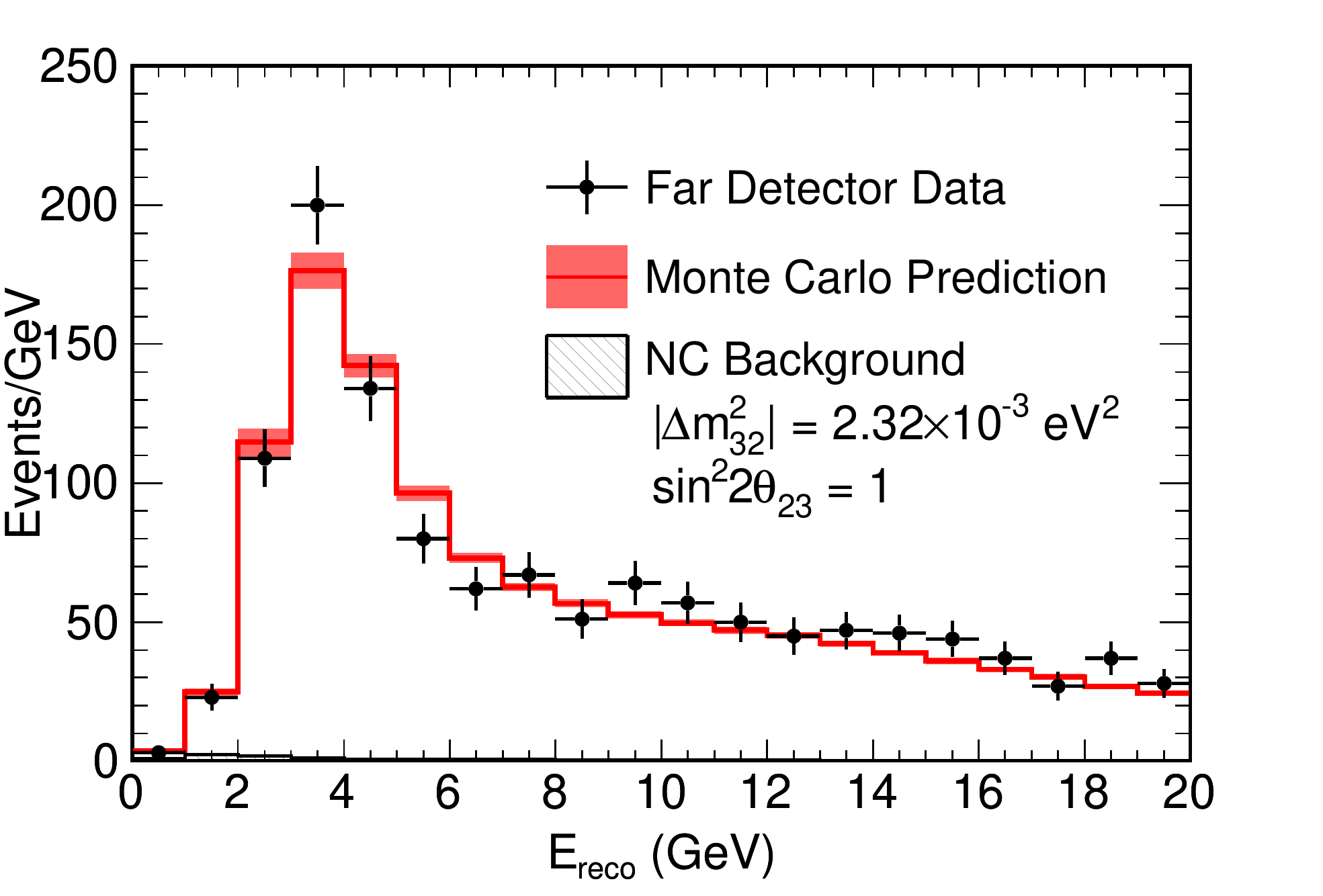}
\includegraphics[height=0.40\textwidth]{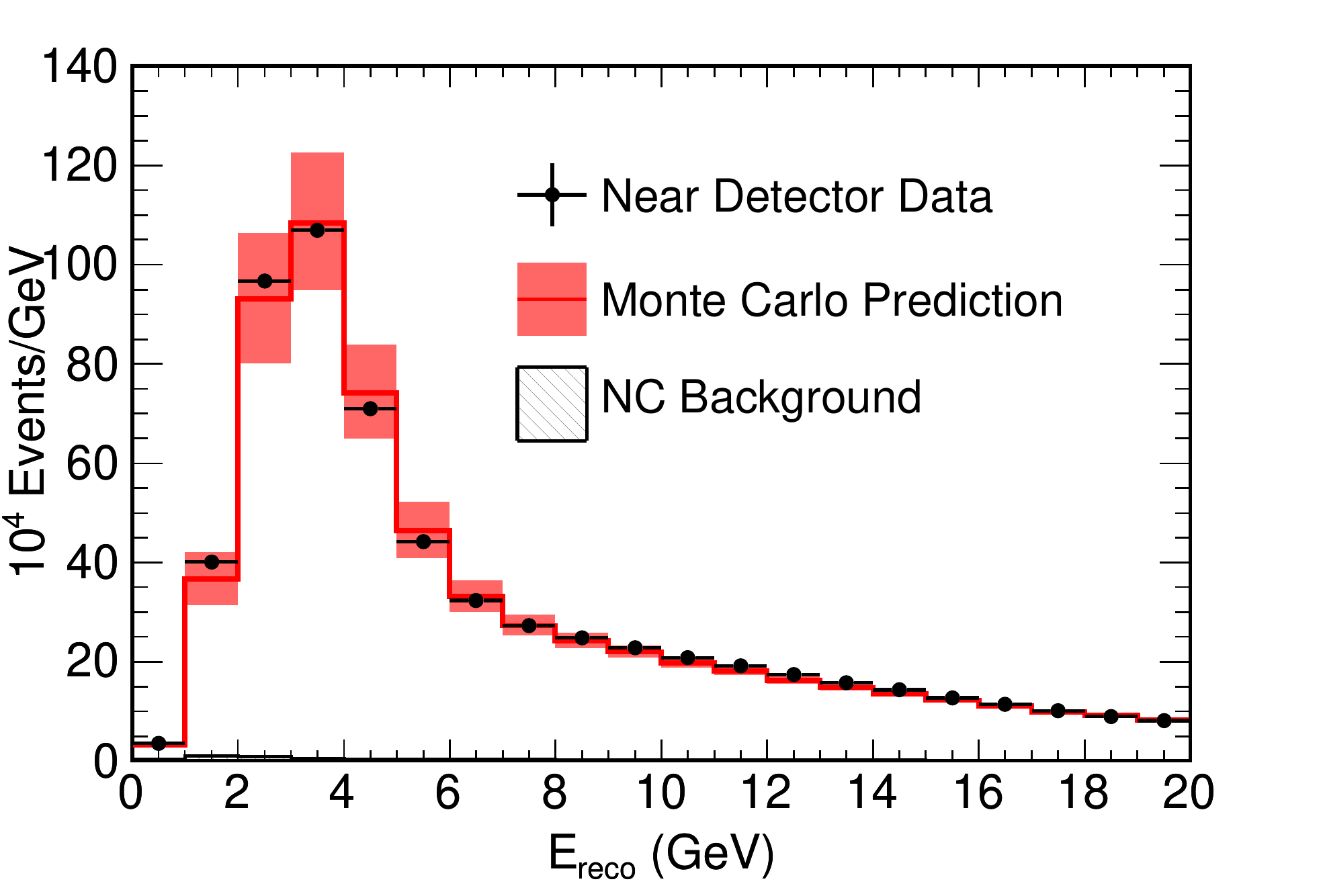}
\caption{Showing the reconstructed energy spectrum for CC \numu interactions in the FD (top) and the ND (bottom). The grey hatched histogram is the NC background and the red line is the three flavour prediction with systematic uncertainties included.}
\label{fig:MINOSsterileCCspec}
\end{figure}

There have been several anomalous results within the neutrino community that have questioned our understanding of neutrinos. The Liquid Scintillator Neutrino Detector (LSND) and MiniBooNE short-baseline experiments saw an excess in the data of electron antineutrinos that can not be explained using the current three-favour model~\cite{ref:LSND,ref:MiniBooNE}; a reinterpretation of reactor fluxes has also lead to a discrepancy in neutrino-oscillation reactor experiments~\cite{ref:ReactorNeutrinoAnomaly}. One explanation is the addition of one or more neutrino types which would oscillate with the three active neutrino flavours; a comprehensive overview of this explanation to account for the above discrepancies can be found in reference~\cite{ref:SterileWhitePaper}.

MINOS has sensitivity to sterile oscillation signatures by looking for perturbations from the three-flavour oscillation formalism in CC events and a deficit in NC events. This MINOS analysis~\cite{ref:MINOSsterile} considers a 3+1 sterile neutrino model. By adding an additional neutrino the PMNS is extended to a $4 \times 4$ matrix which introduces three additional mixing angles $\theta_{24}$, $\theta_{34}$ and $\theta_{14}$. This extra mass state also introduces an extra mass splitting, $\Delta m^{2}_{43}$ where $m_{4} \gg m_{3}$ such that $\Delta m^{2}_{43} \sim \mathcal{O} \left( 1  \,\, \text{eV}^{2}\right)$. An additional deficit of muon neutrinos at the FD aside from the expected loss due to three-flavour oscillations would be an indication of interference from these additional parameters.

\begin{figure}[!ht]
\centering
\includegraphics[height=0.40\textwidth]{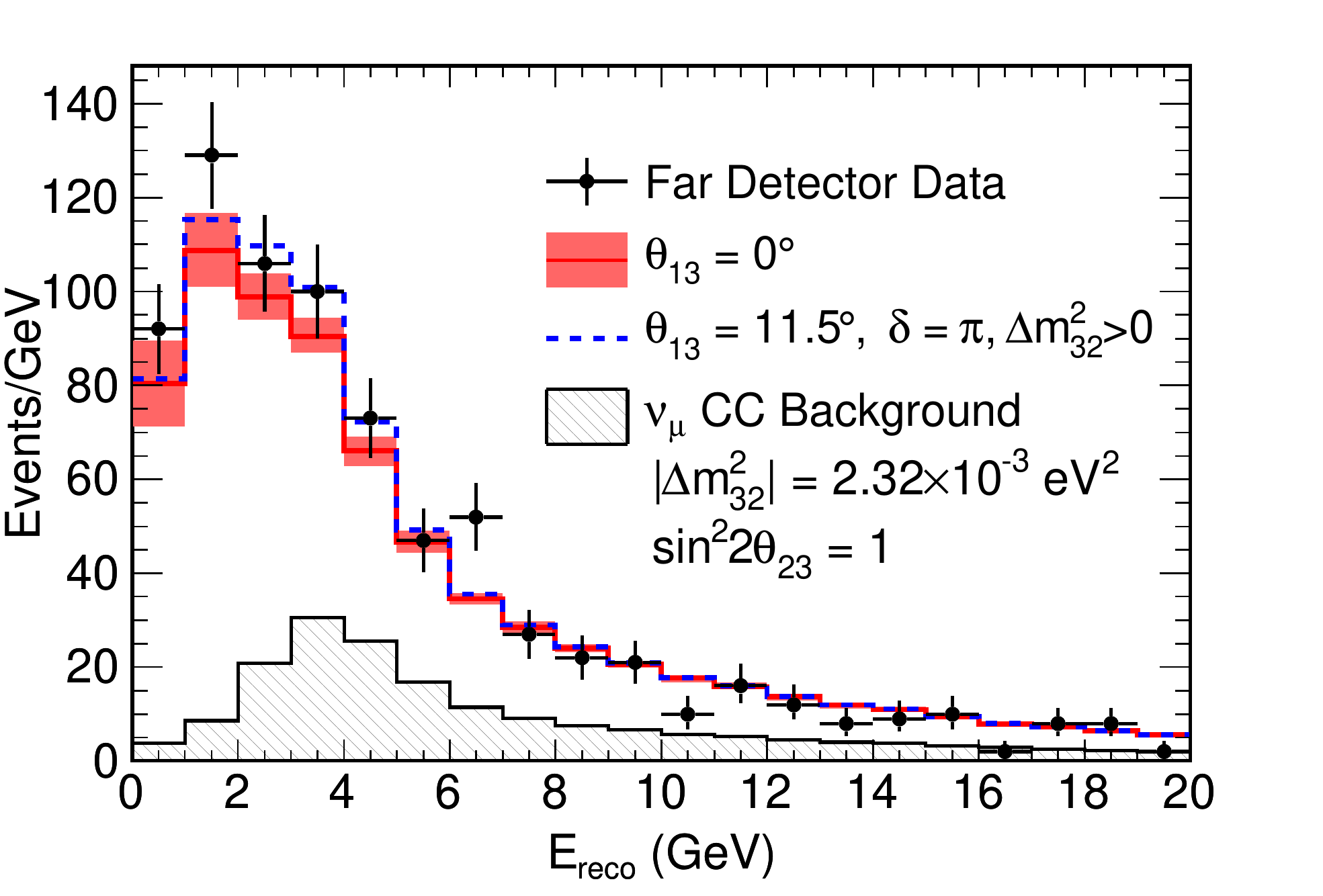}
\includegraphics[height=0.40\textwidth]{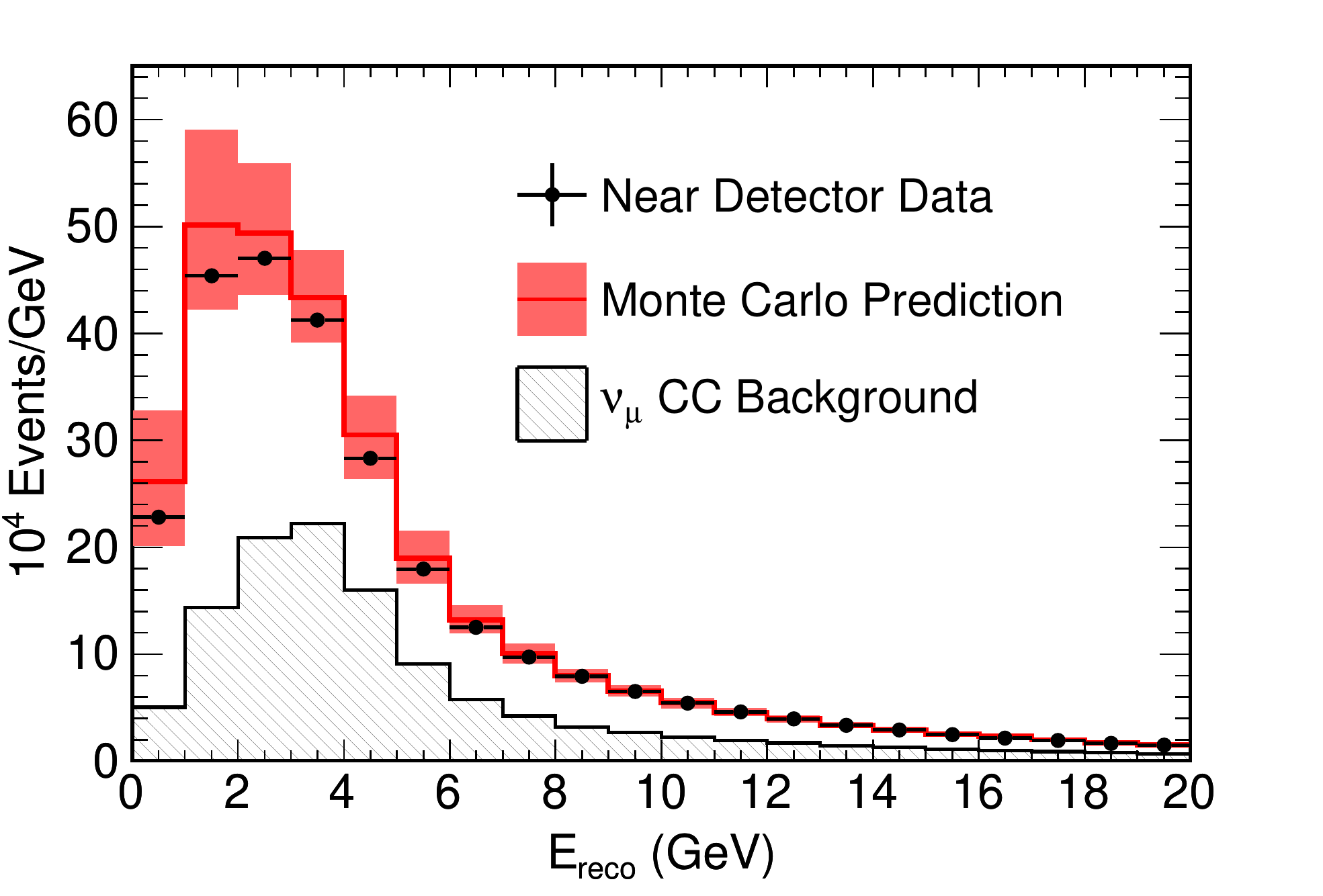}
\caption{Showing the reconstructed energy spectrum for neutral current events at the FD (top) and the ND (bottom). The grey hatched histogram is the \numu CC background and the red line is the three-flavour prediction with systematic uncertainties included. The blue dashed line represents an additional three-flavour prediction with a different $\theta_{13}$ with value $11.5^{o}$}
\label{fig:MINOSsterileNCspec}
\end{figure}

Figures~\ref{fig:MINOSsterileCCspec} and~\ref{fig:MINOSsterileNCspec}  show the NC and CC reconstructed neutrino energy spectrum for the FD and ND and are in good agreement with the expectation of a null sterile neutrino hypothesis. This agreement can be quantified by using the test statistic R for the number of NC events observed at the FD:

\begin{align}
R = \frac{N_{\text{data}} - B_{\text{CC}}}{S_{\text{NC}}},
\label{eqn:Rtest}
\end{align}

\noindent where $N_{\text{data}}$ is the number of events observed, $B_{\text{CC}}$ is the predicted number of CC background interactions, and $S_{\text{NC}}$ is the predicted number of NC interactions in the detector. A value of $R = 1$ would indicate no mixing in the data with sterile neutrinos. The test statistic R is simply based on the integrated number of events, over the full energy range MINOS obtains $R = 1.01 \pm 0.06$ (stat) $\pm 0.05$ (syst) which is in good agreement with the null hypothesis~\cite{ref:MINOSsterile}. 

\begin{figure}[!ht]
\centering
\includegraphics[height=0.60\textwidth]{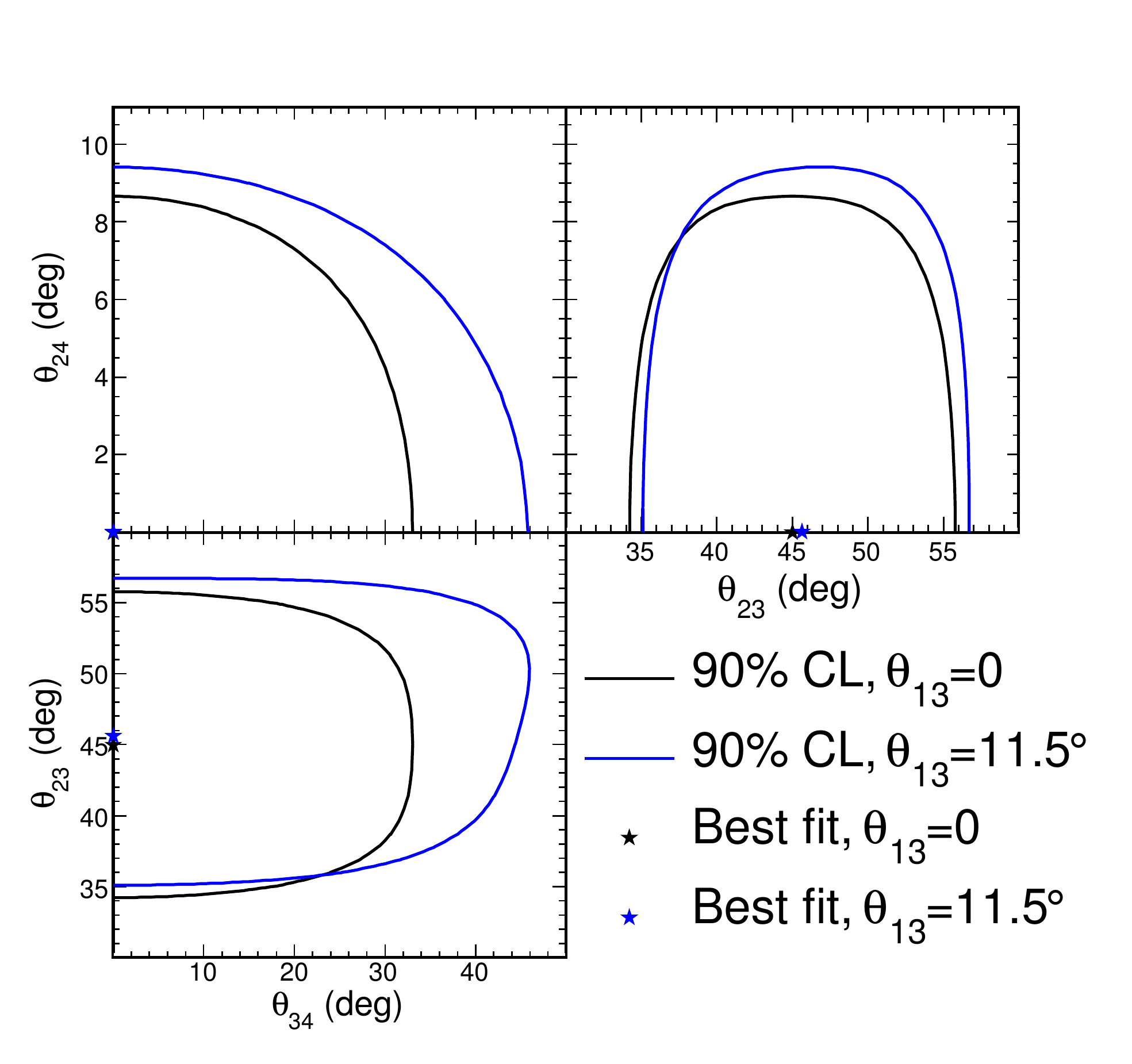}
\caption{Showing the $90\%$ C.L.s for the three mixing angles $\theta_{23}$, $\theta_{34}$, $\theta_{24}$. In the fit the sterile mass splitting was kept fxed at $\Delta m^{2}_{43} = 0.5 \times 10^{-3} \, \text{eV}^{2} $. The black and blue contours are for a different fixed value of $\theta_{13}$ of $0$ ad $11.5^{o}$ respectively.}
\label{fig:sterilecontours}
\end{figure}

MINOS is insensitive to the mixing angle $\theta_{14}$ which is primarily involved in \nue appearance mixing; by looking at muon disappearance MINOS has set $90\%$ C.L.s limits on the other sterile mixing angles yielding $\theta_{24} = \left( 0.0^{+5}_{-0.0} \right)^{\circ}$, and $\theta_{34} = \left( 0.0^{+25}_{-0.0} \right)^{\circ}$~\cite{ref:MINOSsterile}. Figure~\ref{fig:sterilecontours} shows the contours for the mixing angles $\theta_{23}$, $\theta_{34}$, $\theta_{24}$ at a particular value of $\Delta m^{2}_{43} = 0.5 \times 10^{-3} \, \text{eV}^{2} $.

MINOS sets a limit on the sterile-active neutrino coupling by constructing a quantity $f_{s}$, the fraction of \numu that have oscillated into $\nu_{s}$, expressed as:

\begin{align}
f_{s} = \frac{P_{\numu \rightarrow \nu_{s}}}{1 - P_{\numu \rightarrow \numu}}.
\label{eqn:coupling}
\end{align}

For neutrino events around the oscilltion maximum with an energy of 1.4~GeV (the energy for the highest probability of muon neutrino disappearance) a large number of test values are (randomly sampled from Gaussian distributions from the sterile $90\% C.Ls quoted above$) selected for the mixing angles $\theta_{24}$, $\theta_{34}$, and $\theta_{23}$. The value of $f_{s}$ that is larger than 90\% of the test cases is used as the limit, which yields $f_{s} < 0.40$ at a 90\% C.L~\cite{ref:MINOSsterile}.

\section{The Future with MINOS+}

\begin{figure}[!ht]
\centering
\includegraphics[trim={0cm 0cm 0cm 0cm},height=0.50\textwidth, clip]{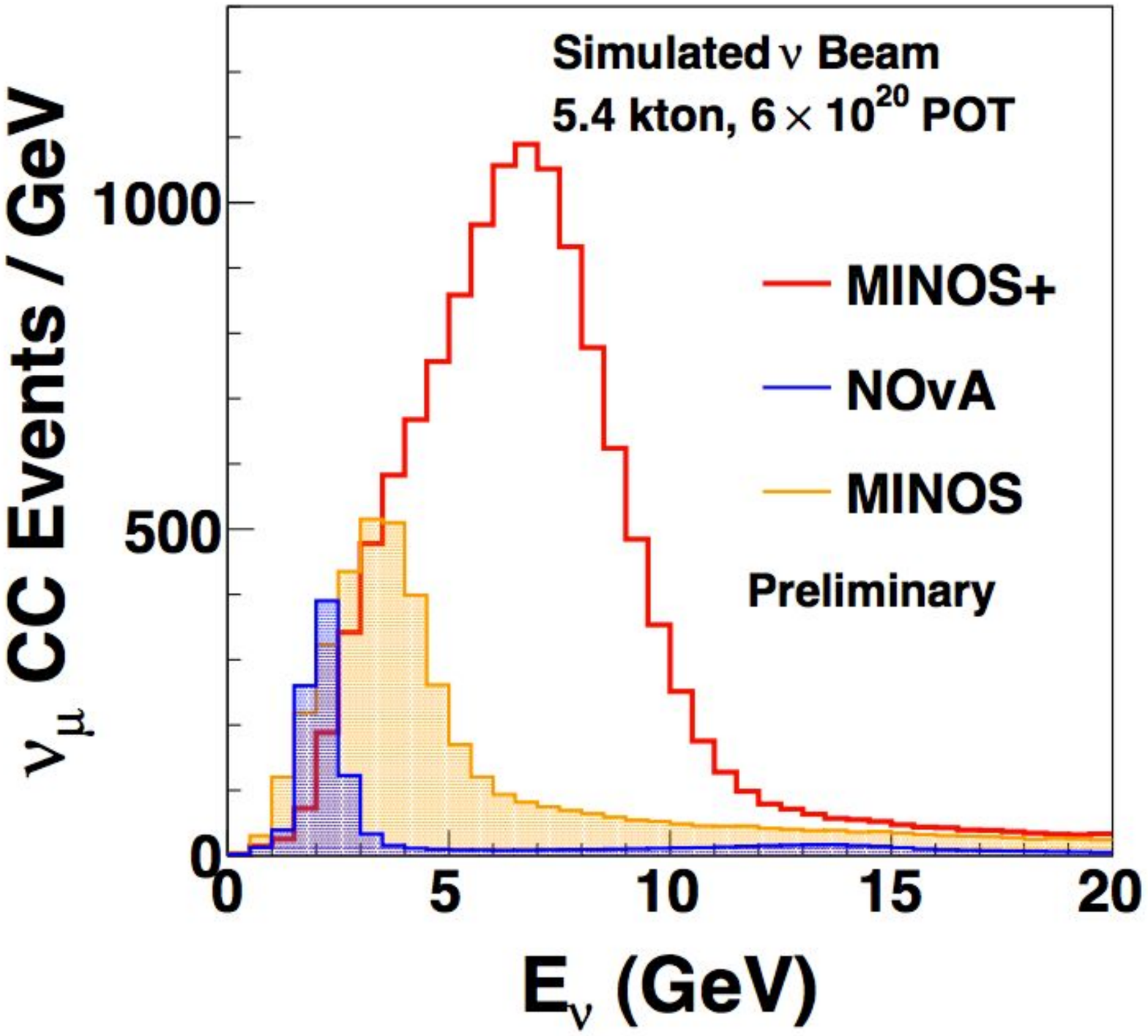}
\caption{The \numu energy spectrum observed for the MINOS, MINOS+ and NO$\nu$A experiments.}
\label{fig:MINOS+spec}
\end{figure}

MINOS+~\cite{ref:MINOSPlusProposal} is the continuation of the MINOS detectors taking data during the NuMI beam run in the medium energy configuration. Being on-axis the neutrino flux at the detectors significantly increases at higher energies as shown in figure~\ref{fig:MINOS+spec}. The beam peak in the medium configuration shifts from 3~GeV to 7~GeV allowing MINOS+ to observe around 4,000 \numu-CC interactions in the FD each year. MINOS+ has been taking data since September 2013, and with the additional statistics will provide a useful contribution to the high-precision test of the three-flavour oscillation formalism and will improve on the world-leading measurements of muon neutrino disappearance made by MINOS. 

\begin{figure}[!ht]
\centering
\includegraphics[trim={0cm 5cm 0cm 4cm}, height=.8\textwidth, clip]{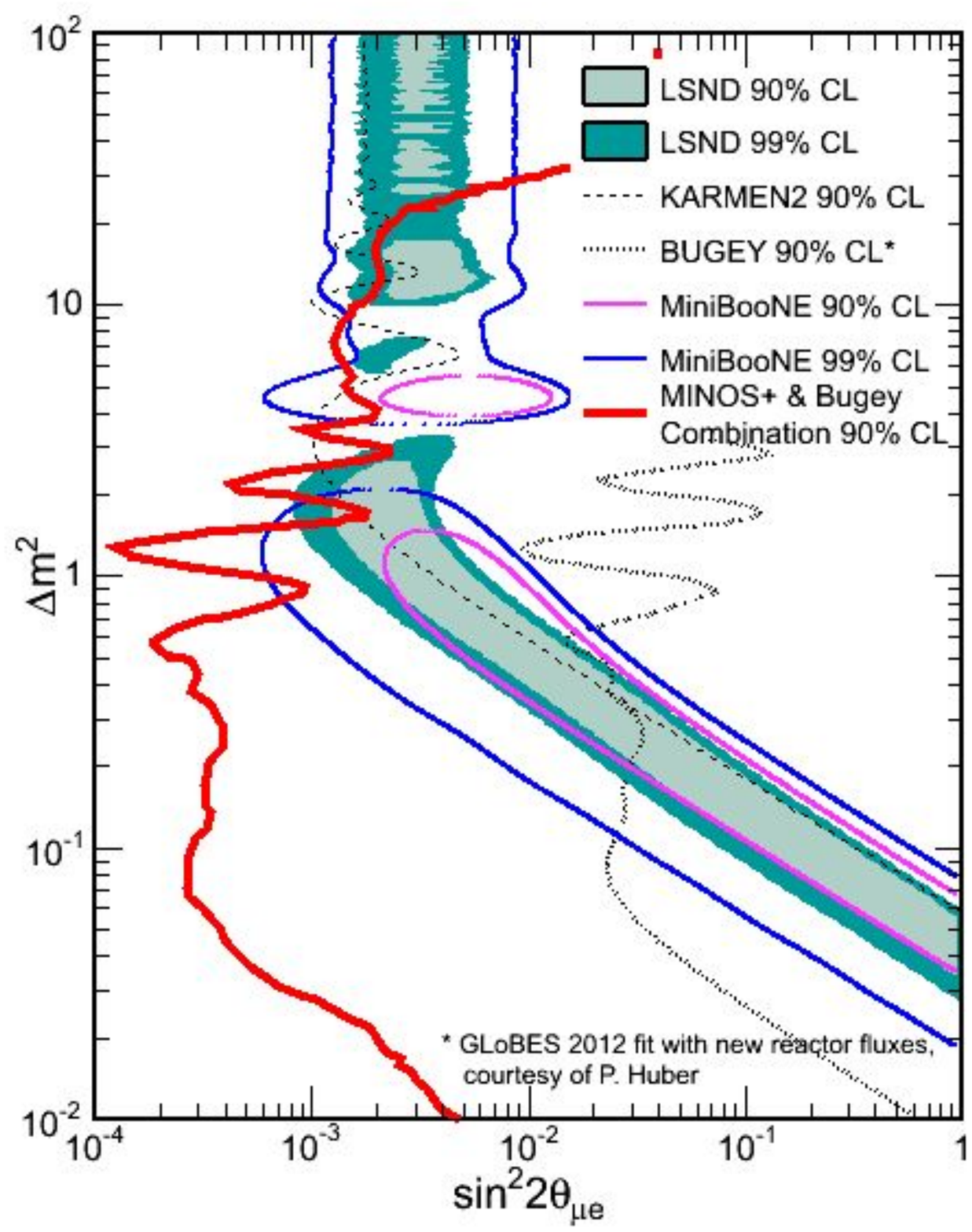}
\caption{The sensitivity of MINOS+ to the existence of sterile
neutrinos, when combined with data from the Bugey~\cite{ref:Bugey} reactor
neutrino experiment. This figure assumes two
years of MINOS+ running with a neutrino-dominated beam.}
\label{fig:bugey}
\end{figure}

With more statistics at high energies MINOS+ is in the unique position to probe and significantly extend the reach of its searches for sterile neutrino signatures in the regions of parameter space favoured by LSND and MiniBooNE. Figure~\ref{fig:bugey} shows a combination between the Bugey~\cite{ref:Bugey} reactor experiment combined with the sensitivity of data taken with MINOS+ assuming two years of MINOS+ running with a neutrino-dominated beam. A combination with an experiment sensitive to $\theta_{14}$ (such as Bugey) is required with the MINOS+ data (sensitive primarily to $\theta_{24}$) in order to set a limit in the LSND style parameter space. The combined 90\% MINOS+-Bugey C.L. excludes a significant amount of the parameter space where sterile neutrinos in a 3+1 model could explain the anomalies seen in past experiments.

\begin{figure}[!ht]
\centering
\includegraphics[height=0.60\textwidth]{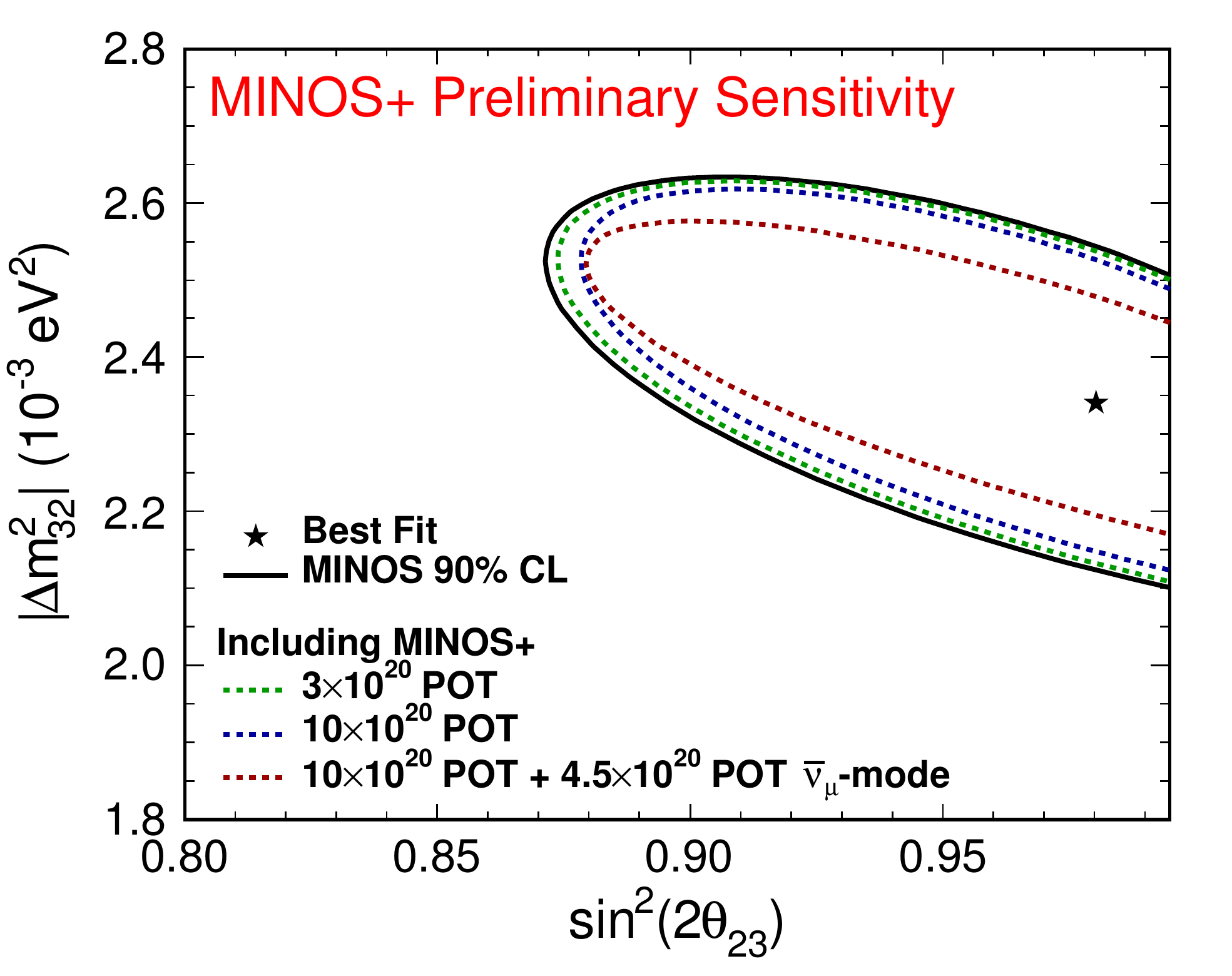}
\caption{Showing the $90 \% $ C.L. contours the MINOS+ experiment can achieve with three different projection running periods. The black contour is the current MINOS result, and the other ones are the MINOS+ sensitivity using simulation.}
\label{fig:NSISensitivity_DelMSqVsSinSq}
\end{figure}

With the increased statistics and higher flux in neutrinos, MINOS+ can also probe non-standard interaction neutrino physics. Figure~\ref{fig:NSISensitivity_DelMSqVsSinSq} shows the increase in sensitivity MINOS+ is to the previous MINOS result with three different possible exposure combinations.

\section{Conclusion}
The MINOS/MINOS+ experiment has been contributing to the neutrino oscillation community for over a decade. In this time the most precise measurement of $\Delta m^{2}_{32}$ has been made. Such a precise measurement is an example of how powerful and necessary a two-detector setup will be for future neutrino oscillation experiments to overcome the large systematics from flux and cross section uncertainties. With the ability to differentiate between neutrinos and anti-neutrinos MINOS has measured oscillation parameters for both and show that they lie in good agreement. Since the discovery of $\theta_{13}$ to be non zero MINOS has been the first experiment to set constraints on the CP violating phase $\delta_{CP}$ as well as incorporating both disappearance and appearance using a full three-flavour framework. MINOS+ will continue taking data and will improve on the MINOS results as well probe at higher energies to investigate the tension in the sterile neutrino debate. 

\acknowledgments

The work of the MINOS and MINOS+ collaborations is supported by the US
DoE, the UK STFC, the US NSF, the State and University of Minnesota,
the University of Athens in Greece, and Brazil's FAPESP and CNPq. We
are grateful to the Minnesota Department of Natural Resources, the
crew of the Soudan Underground Laboratory, and the personnel of
Fermilab, for their vital contributions.

\end{document}